\documentclass[aps,arxiv,twocolumn,superscriptaddress]{revtex4-1}

\usepackage[utf8]{inputenc}
\usepackage[T1]{fontenc}
\usepackage{graphicx}
\usepackage{amsmath}
\usepackage{amssymb}
\usepackage{mathtools}
\usepackage[version=4]{mhchem}

\usepackage{float}

\usepackage{microtype}

\usepackage[colorlinks, allcolors=blue]{hyperref}
\hypersetup{pdftitle={Constructing ANN Potentials},
            pdfauthor={N.~Artrith}}

\makeatletter
\let\Hy@backout\@gobble
\makeatother

\begin{document}

\title{Strategies for the Construction of Machine-Learning Potentials for Accurate and Efficient Atomic-Scale Simulations}

\author{April M. Miksch}%
\email{miksch@theochem.uni-stuttgart.de}
\affiliation{Institute for Theoretical Chemistry, University of Stuttgart, Pfaffenwaldring 55, 70569 Stuttgart, Germany.}%
\author{Tobias Morawietz}%
\affiliation{Bayer AG, Pharmaceuticals, R\&D, Digital Technologies,
  Computational Molecular Design, 42096, Wuppertal, Germany.}%
\author{Johannes Kästner}%
\affiliation{Institute for Theoretical Chemistry, University of Stuttgart, Pfaffenwaldring 55, 70569 Stuttgart, Germany.}%
\author{Alexander Urban}%
\affiliation{Department of Chemical Engineering, Columbia University, New York, New York, 10027, USA.}%
\affiliation{Columbia Center for Computational Electrochemistry, Columbia University, New York, New York, 10027, USA.}%
\author{Nongnuch Artrith$^{*,}$}%
\email{nartrith@atomistic.net}%
\affiliation{Department of Chemical Engineering, Columbia University, New York, New York, 10027, USA.}%
\affiliation{Columbia Center for Computational Electrochemistry, Columbia University, New York, New York, 10027, USA.}%
\date{\today}

\begin{abstract}
Recent advances in machine-learning interatomic potentials have enabled the efficient modeling of complex atomistic systems with an accuracy that is comparable to that of conventional quantum-mechanics based methods.
At the same time, the construction of new machine-learning potentials can seem a daunting task, as it involves data-science techniques that are not yet common in chemistry and materials science.
Here, we provide a tutorial-style overview of strategies and best practices for the construction of artificial neural network (ANN) potentials.
We illustrate the most important aspects of (i)~data collection, (ii)~model selection, (iii)~training and validation, and (iv)~testing and refinement of ANN potentials on the basis of practical examples.
Current research in the areas of active learning and delta learning are also discussed in the context of ANN potentials.
This tutorial review aims at equipping computational chemists and materials scientists with the required background knowledge for ANN potential construction and application, with the intention to accelerate the adoption of the method, so that it can facilitate exciting research that would otherwise be challenging with conventional strategies.
\end{abstract}

\maketitle

{\linespread{1.0}
\tableofcontents
}

\section{Introduction}
\label{sec:introduction}

First-principles-based atomic scale simulations, for example using density-functional theory (DFT)~\cite{hohenberg_inhomogeneous_1964, kohn_self-consistent_1965, burke_perspective_2012, becke_perspective_2014, jones_density_2015, mardirossian_thirty_2017}, can predict many materials properties with quantitative accuracy~\cite{behler_pressure-induced_2008, khaliullin_graphite-diamond_2010, khaliullin_nucleation_2011, sosso_thermal_2012, natarajan_neural_2016, morawietz_how_2016, rowe_development_2018, stricker_machine_2020}.
However, they are usually limited to small atomic structures with less than 1,000~atoms and time scales on the order of nanoseconds, owing to the high computational demand and polynomial scaling with the system size.
During the last decade, a family of methods for accelerating first-principles sampling based on machine learning (ML) has emerged~\cite{behler2016perspective, artrith_machine_2019, mueller2020machine, noe2020machine, unke_high-dimensional_2020, morawietz_machine_2020, unke2021machine}, which holds promise to overcome these limitations.
ML regression models, usually based on artificial neural networks (ANN)~\cite{lorenz2004,behler_generalized_2007} or Gaussian process regression (GPR)~\cite{bartok_gaussian_2010}, are trained to interpolate the outcomes from first-principles calculations, so that the trained ML model can be used as computationally efficient drop-in replacements for the original method.

Predicting the preferred atomic structure at specific thermodynamic conditions and its evolution over a certain period of time requires a description of the relative energy of atomic arrangements, i.e., the potential energy surface (PES).
First-principles methods, such as DFT or quantum-chemical methods based on wavefunction theory~\cite{moller_note_1934, coester_short-range_1960, cizek_correlation_1966, bartlett_coupled-cluster_2007, cremer_moller-plesset_2011, zhang_coupled_2019}, approximate the PES based on the interactions of electrons and atomic nuclei arising from the laws of quantum mechanics.

Most ML potential (MLP) approaches do not consider electronic interactions explicitly but instead approximate the PES as a function of the atomic positions only.
For many modeling applications, this general strategy allows MLPs to deliver the accuracy of the reference method at a computational cost that is orders of magnitude lower and scales only linearly with the number of atoms.
Owing to the success of early ANN-based Behler-Parrinello MLPs~\cite{behler_generalized_2007, behler_first_2017} and GPR-based Gaussian approximation potentials (GAP)~\cite{bartok_gaussian_2010, bartok_representing_2013}, the number of MLP methods proposed in the literature has been rapidly growing:
Examples include MLPs based on GPR and other kernel-based methods, such as kernel ridge regression~\cite{botu_adaptive_2015, huan_universal_2017, john_many-body_2017, nyshadham_machine-learned_2019}, moment tensor potentials~\cite{shapeev_moment_2016, novikov_MLIP_2021}, graph-networks using message passing~\cite{pmlr-v70-gilmer17a, duvenaud_convolutional_2015, kearnes_molecular_2016, schutt_quantum-chemical_2017, chen_graph_2018, jorgensen_neural_2018, schutt_schnet_2018, unke_physnet_2019, xie_crystal_2018}, spectral neighbor analysis potentials (SNAP)~\cite{thompson_spectral_2015, wood_extending_2018}, and other ANN-based approaches~\cite{artrith_implementation_2016, smith_ani-1_2017, zhang_deep_2018, westermayr2019machine}.
We emphasize that this list is not exhaustive and also does not include the various ML methods for atomistic modeling that cannot be considered interatomic potentials~\cite{tibshirani_regression_1996, hastie_elements_2009, mueller_bayesian_2009, brown_efficient_2010, balabin_support_2011, hansen_assessment_2013, li_molecular_2015, seko_first-principles_2015, chmiela_machine_2017, gastegger_machine_2017, butler_machine_2018, cao_use_2018, hy_predicting_2018, mardt_vampnets_2018, ryczko_convolutional_2018, artrith_learning_2020, artrith_predicting_2020}.
For a comparison of different MLP methods we refer the reader also to perspectives and reviews by Behler~\cite{behler2016perspective}, Deringer~\cite{deringer2019machine}, Mueller~\cite{mueller2020machine}, No\'e~\cite{noe2020machine}, and Unke~\cite{unke_high-dimensional_2020, unke2021machine}.

Although the ML regression models can be used in simulations in the same fashion as conventional interatomic potentials (\emph{force fields})~\cite{jorgensen_potential_2005, becker_considerations_2013}, the construction and applicability range of ML potentials is significantly different.
Here, we review and discuss the practical aspects of constructing and validating MLPs in the form of a tutorial with concrete examples.
To be as specific as possible, the tutorial focuses on ANN-based MLPs (ANN potentials), although many aspects of the discussion on data selection (section~\ref{sec:data}) and active learning (section~\ref{sec:testing-and-refinement}) apply to other MLP methods as well.
The sections on model selection (section~\ref{sec:model}) and training/validation (section~\ref{training}) are mostly specific to ANN potentials.

\begin{figure*}[t]
  \centering
  \includegraphics[width=0.8\textwidth]{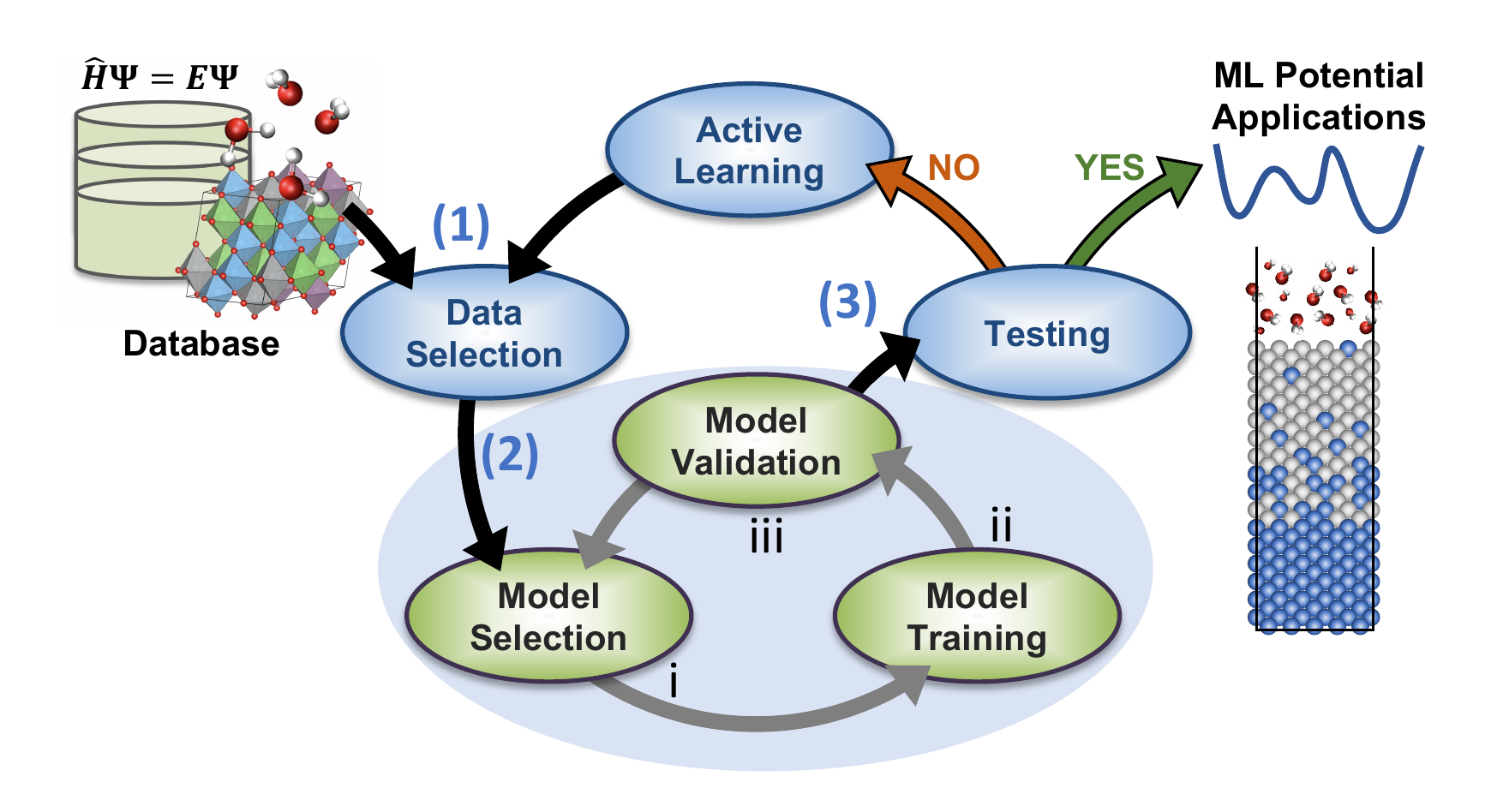}
  \caption{\label{fig:TOC-figure}%
    \textbf{Iterative construction of a machine-learning potential (MLP) based on artificial neural networks (ANNs).}  The process starts with \textbf{(1)} an initial data set, which is then used for \textbf{(2)} model construction, i.e, a model is selected, trained, and its hyperparameters are validated.  The accuracy of the trained model is then assessed in \textbf{(3)} a testing step.  If the MLP passes the testing, it is ready for applications.  Otherwise, additional data points are included in the reference data set (generated through active learning), and the process is repeated.}
\end{figure*}

The construction of ANN potentials and other MLPs is centered around the compilation of reference data sets with atomic structures and their corresponding first principles energies, and potentially further information such as interatomic forces and atomic charges.
Unlike many conventional potentials with functional forms derived from physical approximations, MLPs are usually not capable of extrapolating to atomic structures and compositions that lie outside of the PES region described by the reference data.
The reference data set therefore needs to span the entire structural and chemical space that the MLP must represent for the intended application range, while it should include as few unnecessary or redundant data points as possible.
In practice, MLP construction and the compilation of reference data is therefore (typically) an iterative process, shown in \textbf{Figure~\ref{fig:TOC-figure}}, that involves:
\begin{enumerate}
\item Data collection,
\item Model construction, and
\item Testing,
\end{enumerate}
which is repeated until the MLP passes the testing step with the desired accuracy.
In each iteration, the reference data set is extended through an \emph{active learning} scheme.

As also hinted at in \textbf{Figure~\ref{fig:TOC-figure}}, the model construction consists itself of three iterative steps:
(i)~\emph{Model selection} is the process of deciding the type, \emph{complexity}, and other \emph{hyperparameters} of the ML model;
(ii)~\emph{training} is the optimization of the adjustable model parameters to best reproduce the reference data set, as measured by a \emph{loss function}; and
(iii)~in the \emph{validation} step, \emph{over- or underfitting} is detected, and, if necessary, the process is repeated from step~(i) with adjusted model complexity and hyperparameters.

In this work, we discuss each of the steps outlined in \textbf{Figure~\ref{fig:TOC-figure}} from a practical perspective.
The next section first provides a brief introduction to the ANN potential method, which is followed by separate sections on data selection, the individual steps of model construction, testing, and active learning.
In the final discussion section, current limitations and advanced techniques of the ANN potential method are considered.

\subsection{High-dimensional artificial neural network potentials}
\label{sec:ANN-potentials}

\begin{figure*}[t]
  \centering
  \includegraphics[width=\textwidth]{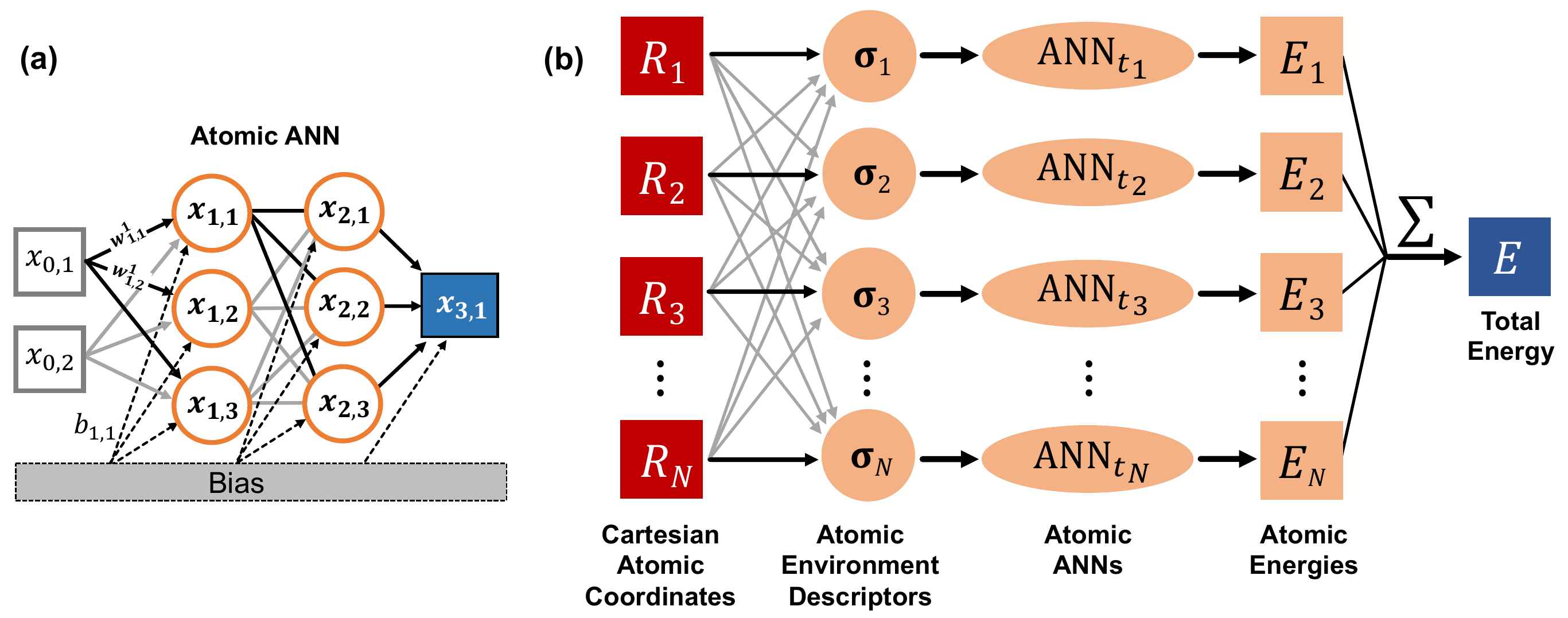}
  \caption{\label{fig:high-dim-ANN-potentials}%
  \textbf{(a)}~Graph representation of an example feedforward artificial neural network (ANN) with two input nodes ($x_{0,1}$ and $x_{0,2}$), one output node ($x_{3,1}$), and two hidden layers with each three nodes.  The operation performed by each node is given in equation~\eqref{eq:artificial-neuron}.  In an ANN potential, the input nodes correspond to features of an atomic environment $\sigma_{i}$ and the value of the output node is equal to an atomic energy $E_{i}$. \textbf{(b)}~Diagram of the high-dimensional neural network that combines the atomic ANNs of all atoms in a structure for an $N$-atom system. The output is the total energy $E$, which is the sum of the individual atomic energy contributions $E_i$, which are in turn the outputs of atomic feedforward ANNs as shown in panel (a).}
\end{figure*}

ANNs are a class of mathematical functions that can be represented by graphs which resemble networks of nodes (\emph{artificial neurons}) that calculate the weighted sum of multiple input values and apply an \emph{activation function} to the result.
The operation performed by a single node can be expressed as
\begin{align}
  x_{i,j}
  = f^{i}_{\textup{a}}\Biggl(
      \sum_{k}^{N_{i-1}} w^{i}_{k,j} x_{i-1,k} + b_{i,j}
  \Biggr)
  \label{eq:artificial-neuron}
\end{align}
where the output $x_{i,j}$ is the value of the $j$-th node in the $i$-th layer of the ANN, the input value $x_{i-1,k}$ is the $k$-th node in the previous layer $(i-1)$, $w^{i}_{k,j}$ is the \emph{weight} of the input value, $b_{i,j}$ is an additional \emph{bias weight} that is always added to the input irrespective of the input values, and $f^{i}_{\textup{a}}$ is an \emph{activation} (or \emph{transfer}) function.
Equivalently, equation~\eqref{eq:artificial-neuron} can also be expressed using matrix-vector operations
\begin{align}
  \vec{x}_{i}
  = f^{i}_{\textup{a}}\Bigl(
    \mathbf{W}_{i} \vec{x}_{i-1} + \vec{b}_{i}
  \Bigr)
  \quad ,
  \label{eq:feedforward-layer}
\end{align}
where the vectors $\vec{x}_{i-1}=(x_{i-1,1}, \ldots, x_{i-1,N_{i-1}})^{T}$ and $\vec{x}_{i}=(x_{i,1}, \ldots, x_{i,N_{i}})^{T}$ contain the values of all nodes in the input and output \emph{layers}, respectively, $(\mathbf{W}_{i})_{k,j} = w^{i}_{k,j}$ is the weight matrix, and  $(\vec{b}_{i}) = b_{i,j}$.
ANN potentials are based on \emph{feedforward} ANNs in which layers of nodes are connected such that values are passed layer-by-layer from an input layer through one or more hidden layers to an output layer.
A graph representation of an example feedforward ANN is shown in \textbf{Figure~\ref{fig:high-dim-ANN-potentials}a}.
The number of layers, the number of nodes per layer, and the choice of activation function are hyperparameters that need to be chosen in the model selection step and are discussed in section~\ref{sec:ANN-parameters}.
Note that the input dimension (number of nodes in the input layer) and output dimension (number of nodes in output layer), i.e., the dimensions of the domain and co-domain of the ANN function, are fixed for a given ANN.
ANN training is the process of optimizing the weight parameters $\{w^{i}_{k,j}\}$ and $\{b_{i,j}\}$ such that the target values of the reference data set are reproduced as well as possible (see section~\ref{training}).
For a more thorough introduction to ANNs, we refer the reader to previous literature~\cite{artrith_implementation_2016} and standard textbooks~\cite{montavon_tricks_2012}.

In principle, PESs can be directly represented by ANNs, feeding the atomic coordinates (in form of internal coordinates) into the input layer and producing the potential energy as the sole value of the output layer~\cite{blank_neural_1995}.
While useful for many applications, such direct ANN-PES models are limited to a fixed number of atoms and do not automatically reflect the physical invariances of the potential energy with respect to rotation and translation of the entire atomic structure and the exchange of equivalent atoms.
As such, direct ANN-PES models are not comparable to conventional interatomic potentials.

To leverage the flexibility of ANNs for the construction of actual, reusable interatomic potentials, Behler and Parrinello proposed an alternative approach~\cite{behler_generalized_2007} that was later extended to multiple chemical species by Artrith, Morawietz, and Behler~\cite{artrith_high-dimensional_2011}, in which the total energy $E(\sigma)$ of an atomic structure $\sigma$ is decomposed into atomic energy contributions $E_{i}$
\begin{align}
\begin{aligned}
  E(\sigma) &\approx \sum_{i}^{\textup{atoms}} E_{i}(\sigma_{i})
  \\
  \quad\text{with}\quad
  \sigma_{i} &= \{ \vec{R}_{j}, t_{j} \;\text{for}\;
     |\vec{R}_{j} - \vec{R}_{i}|\leq R_{\textup{cut}}\}
\end{aligned}
\label{eq:E=sum-E_i}
\quad ,
\end{align}
where $\sigma_{i}$ represents the \emph{local atomic environment} of atom $i$ that contains the positions $\vec{R}_{j}$ and species $t_{j}$ of all atoms within a cutoff distance $R_{\textup{cut}}$ from atom $i$, including the position $\vec{R}_{i}$ and type $t_{i}$ of atom $i$ itself.
In the high-dimensional ANN potential method by Behler and Parrinello, the atomic energies $E_{i}(\sigma_{i})$ are predicted by species-specific ANNs.
A graph representation of such a high-dimensional ANN potential is shown in \textbf{Figure~\ref{fig:high-dim-ANN-potentials}b}.

As expressed in equation~\eqref{eq:E=sum-E_i}, the ANN potential does not yet incorporate the invariances of the potential energy, since the input of the atomic ANNs are still atomic (Cartesian) coordinates.
Even worse, the number of atoms within the local atomic environment is generally structure dependent, but the input dimension of ANNs is fixed, which would render the atomic ANNs essentially not transferable to other structures.
These limitations can be removed by representing the atomic coordinates within local atomic environments with a fixed number of \emph{features} that have the same invariances as the potential energy.
Once the atomic species are also encoded, a \emph{fingerprint} $\widetilde{\sigma}_{i}$ of the local atomic environment $\sigma_{i}$ is obtained that can be used as input for the atomic ANNs, so that equation~\eqref{eq:E=sum-E_i} can be written as
\begin{align}
  E(\sigma)
  \approx E_{\textsc{ann}}(\sigma)
  = \sum_{t}^{\substack{\textup{atom}\\\textup{types}}}
    \sum_{i}^{\substack{\textup{atoms of}\\\textup{type $t$}}}
    \textup{ANN}_{t}(\widetilde{\sigma}_{i})
  \label{eq:multicomponent-HDNNP}
  \quad ,
\end{align}
where $\textup{ANN}_{t}$ is the atomic ANN for atoms of type $t$.

It is important to note that the high-dimensional ANN potential method
as introduced by Behler and Parrinello uses ANNs to represent atomic energies, even though \emph{the training
  targets are the total energies}.
Atomic energies are not uniquely defined and are not a quantum-mechanical observable.
ANN potential training, i.e., the weight optimization problem, can be expressed completely in terms of the total energies without knowledge of the \emph{atomic energies}, as discussed further in section~\ref{sec:loss_function}.
It is possible to come up with rigorous (but non-unique) definitions of the atomic energies and use those as targets for training~\cite{huang2019}, but such schemes require additional processing of first principles calculations and are not considered in the following.

Further note that the atomic-energy approach of equation~\eqref{eq:multicomponent-HDNNP} can only work with good accuracy if the total energy is determined fully by short-ranged interactions.
Long- and intermediate-ranged interactions, such as electrostatic (\emph{Coulomb}) and dispersive (\emph{London} or \emph{van der Waals}) interactions need to be accounted for separately.
Extensions of the high-dimensional ANN potential method to electrostatic and dispersive interactions have been developed and are briefly discussed in section~\ref{sec:discussion}.

\section{Data Selection}
\label{sec:data}

Since the mathematical form of ANN potentials and other types of MLPs is unconstrained and not derived from physical approximations, MLPs are poor at extrapolating to atomic structures or compositions that are very different from those included in the data that the MLP was trained on.
The lack of extrapolation capabilities is, in fact, a general property of ANNs~\cite{haley_extrapolation_1992}.
The quality of an ANN potential therefore depends strongly on the reference data set that it is trained on:
\begin{itemize}
\item An MLP's \emph{accuracy} for the prediction of materials or molecular properties cannot exceed that of the reference method, and
\item The \emph{transferability} (the ability to \emph{generalize}) of an MLP is determined by the structural and chemical space that is represented in the reference data set.
\end{itemize}
Incorrect data points, noise, and redundant data can further impede the MLP training process.
Data selection is therefore of great importance for the construction of accurate and transferable MLPs~\cite{smith_less_2018,loeffler_active_2020}.

The atomic structures and compositions within the reference data set determine the \emph{feature space}. The \emph{target space} depends on the types of derived physical properties that are present in the data set and can be represented by the selected model (see section~\ref{sec:model}).
In addition to total energies, other quantifiable physical properties, such as atomic charges~\cite{artrith_high-dimensional_2011, grisafi_incorporating_2019, unke_physnet_2019}, electronegativities~\cite{ghasemi_interatomic_2015, faraji_high_2017, ko_fourth-generation_2020}, molecular dipole moments~\cite{gastegger_machine_2017, litman_temperature_2020}, and atomic forces or higher-order derivatives~\cite{artrith_high-dimensional_2011, cooper_potential_2018, singraber_parallel_2019} can in principle be included in the reference data set.

In this section, we discuss strategies for the compilation of \emph{initial} reference data sets, i.e., data sets generated before any model construction or testing has occurred.
The refinement of reference data in subsequent iterations of the process in \textbf{Figure~\ref{fig:TOC-figure}} is discussed in the context of active learning approaches in section~\ref{sec:testing-and-refinement}.

While data selection is crucial for any kind of MLP, there are some specifics that apply to ANN potentials:
In their conventional form, ANN potentials can be most easily trained on energies and forces (see also section~\ref{training}), but it is challenging to include other target properties of interest (such as higher order derivatives of the potential energy) as well.
This is an area where other MLP methods could be better suited.
The computational cost of evaluating ANN potentials does not increase with the size of the reference data set, so that ANN potentials can be trained on very large reference data sets containing millions of data points~\cite{smith_ani-1_2017}.
This is not generally the case for MLP methods, and depending on the method it can be beneficial to reduce the size of the reference data set.

\subsection{Recipe: Generation of initial reference data sets}
\label{sec:recipe-initial-data}

The initial reference data set is used to \emph{kick off} the iterative refinement of the MLP shown in the flow chart of \textbf{Figure~\ref{fig:TOC-figure}}.
The data set should already include structures that are close to those relevant for the eventual application of the potential.
An initial data set could, for example, comprise of
\begin{enumerate}
\item Relevant ideal atomic structures from databases, e.g., ideal crystal structures from crystallographic databases;
\item Structures that were \emph{derived} from the ideal structures by modifying the \emph{atomic positions}, e.g., by displacing atoms;
\item Derived structures that were generated by altering the \emph{lattice parameters} or by scaling the ideal structures; and
\item Derived structures that exhibit relevant defects, e.g., point defects, substitutional disorder, etc.
\end{enumerate}
Including distorted structures is particularly important, so that the initial data set contains structures with bond length that are significantly shorter and longer than those in the ideal structures.
While compiling the initial data set, it is important to keep an eye on its energy distribution: (i)~structures that are so high in relative energy that they are realistically never encountered in the planned simulations should not be included in the data set, (ii)~the data set should not exhibit gaps in energy space, i.e., the energies between the most and least stable structures should be relatively evenly represented in the data set.
Note that the maximal reasonable relative energy depends on the specific application that the potential is going to be used for.

\begin{figure*}[t]
    \centering
\includegraphics[width=0.9\textwidth]{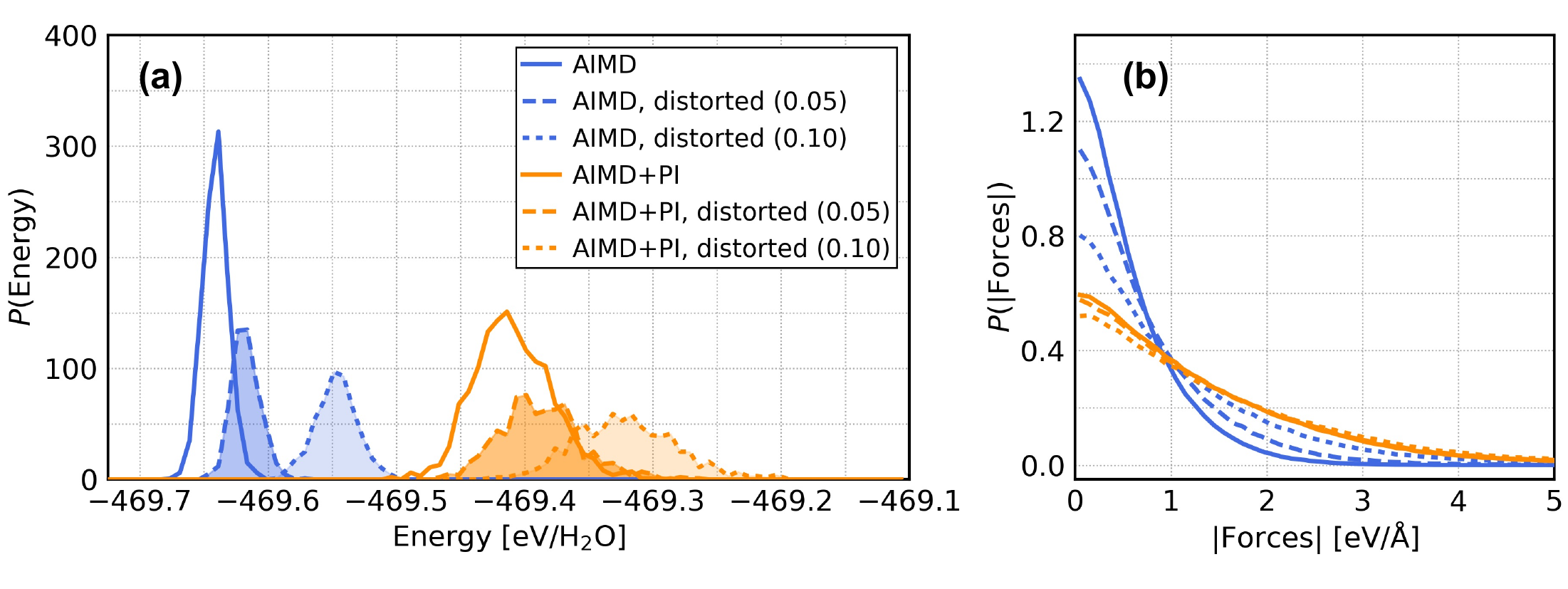}
    \caption{Distribution of \textbf{(a)} total energies and \textbf{(b)} forces of the reference configurations used to train the initial machine learning potential (MLP) for liquid water (64 \ce{H2O} molecules)~\cite{morawietz_interplay_2018}. Reference configurations were obtained from \textit{ab initio} molecular dynamics (AIMD)  trajectories with classical and quantum nuclei (AIMD+PI)~\cite{marsalek_quantum_2017}. In addition to the MD snapshots, distorted configurations with higher forces where added by randomly displacing Cartesian coordinates by a maximum displacement of 0.05 and 0.10 Angstrom, respectively.}
    \label{fig:water-initial-data}
\end{figure*}

\subsection{Example: An initial reference data set for liquid water}
\label{sec:example-initial-data}

MLP-based simulation have previously been shown to achieve high accuracy for both liquid and solid water~\cite{morawietz_densityfunctional_2013, morawietz_how_2016, morawietz_interplay_2018, morawietz_hiding_2019, cheng2019water, loeffler_active_2020}.
Here, we illustrate some of the considerations that go into the selection of an initial data set using the example of an MLP for liquid water which was applied to the calculation of vibrational spectra across the full liquid temperature range~\cite{morawietz_interplay_2018, morawietz_hiding_2019}.
\textbf{Figure~\ref{fig:water-initial-data}a~and~b} show the distribution of interatomic energies and forces in an initial reference data set for the liquid water MLP.
In this particular case, the data set was obtained by selecting periodic water structures (containing 64~\ce{H2O} molecules) from DFT-based \emph{ab initio} molecular dynamics (AIMD) simulations at ambient conditions, in which the thermal motion of the atoms was used to sample the relevant structure space~\cite{marsalek_quantum_2017}.
In addition to configurations from AIMD simulations with classical nuclei, the initial data set also contains snapshots from AIMD trajectories with \emph{quantum nuclei} as described by the path integral (PI) method~\cite{markland_nuclear_2018} which are located in a higher energy region of the PES (shown in orange in \textbf{Figure~\ref{fig:water-initial-data}}).
Furthermore, distorted structures were generated by randomly displacing atoms from the original structures (obtained from equally spaced snapshots of the AIMD and AIMD+PI trajectories) with maximum displacements of 0.05~\AA{} and 0.10~\AA{}, respectively, to further increase the structural diversity in the initial reference data set.
The initial data set comprising of a total of 5,369 atomic structures was then used for the construction of an initial ANN potential.
As seen in \textbf{Figure~\ref{fig:water-initial-data}}, the initial structures fully cover a large energy range of 500~meV/\ce{H2O} (the thermal energy $k_{\textsc{b}}T$, where $k_{\textsc{b}}$ is Boltzmann's constant, at 300~K is around 26~meV per degree of freedom) which can be expected to be sufficient for performing stable MD simulations even at temperatures above ambient conditions with the ANN potential trained to this initial data set.

\section{Model Selection}
\label{sec:model}

Once an initial data set has been compiled, the next step in the flow chart of \textbf{Figure~\ref{fig:TOC-figure}} is the selection of a model to represent the data.
This means, in the case of ANN potentials, to decide on \textbf{(A)} the descriptor that is used for the featurization of local atomic environments, i.e., the \emph{fingerprint} $\widetilde{\sigma}_{i}$ in equation~\eqref{eq:multicomponent-HDNNP}, and
\textbf{(B)} the ANN hyperparameters, for example, the number of ANN layers, the number of nodes per layer, and the type of activation function.

\subsection{Recipe: Featurization of local atomic environments}
\label{sec:featurization}

As discussed in section~\ref{sec:ANN-potentials}, to be suitable as input for an ANN, the coordinates and atomic species of the atoms within the local atomic environment $\sigma_{i}$ of an atom $i$ need to be transformed into a feature vector with fixed length that should also be invariant with respect to rotation and translation of the atomic structure and the exchange of equivalent atoms.
The amount of detail captured by this \emph{descriptor} also determines the ability of the ANN potential to distinguish between different atomic structures:
If the descriptor is too approximate, different atomic structures might yield the same feature vector.
Hence, the choice of descriptor is crucial for the accuracy of the ANN potential.

The choices to be made are (1)~the type of descriptor that is used for the featurization of the local atomic structure and compositions, (2)~the resolution of the descriptor, and (3)~the radial cutoff, i.e., the size of the local atomic environment.

\subsubsection{Selecting a descriptor}
\label{sec:descriptor-type}

Various descriptors have been proposed in the literature~\cite{prsaeifard_assessment_2020}, most of which are derived from basis-set expansions of either the atomic density of the local atomic environment~\cite{bartok_gaussian_2010, bartok_representing_2013, khorshidi_amp_2016, unke_reactive_2018, kocer_novel_2019, zaverkin_gaussian_2020}, the radial and angular distribution functions (RDF and ADF) and higher-order distribution functions within the local atomic environment~\cite{behler_generalized_2007, behler_atom-centered_2011, artrith_efficient_2017, smith_ani-1_2017, tjocp148-2018-241709, faber_alchemical_2018, christensen_fchl_2020}, or directly the local potential energy surface~\cite{shapeev_moment_2016, gubaev_machine_2018}.
In addition to differences in the features for the geometry of the local atomic environment, the above descriptors also differ in their approach for encoding chemical species.
We note that the above list of descriptors is not exhaustive, and the development of new methods is currently an active field of research.
Several of the expansion-based descriptors are available in open-source libraries~\cite{reveil_classification_2018, himanen_dscribe_2020}.
As an alternative to expansion-based descriptors for \emph{hand-crafted} feature construction, an invariant representation of the atomic coordinates can also be machine learned~\cite{schutt_quantum-chemical_2017, schutt_schnet_2018, lubbers_hierarchical_2018, schutt_schnetpack_2019, unke_physnet_2019, nikitin_dracon_2020}.

For the purpose of ANN potentials, an ideal descriptor
\begin{enumerate}
\item Exhibits the symmetries of the potential energy,
\item Requires minimal manual fine-tuning,
\item Provides a parameter for adjusting its resolution,
\item Is continuous and differentiable, so that analytical forces can be obtained,
\item Is computationally efficient, and
\item Does not scale with the number of chemical species.
\end{enumerate}
Here, we discuss one specific choice of descriptor that fulfills the above criteria, has been frequently used for the construction of ANN potentials~\cite{lacivita_structural_2018, sun_fast_2019, cooper_efficient_2020, chen_exploiting_2020, mori_neural_2020}, and is available in the free and open-source ANN potential package \ae{}net~\cite{artrith_implementation_2016}.

\begin{figure*}[t]
  \centering
  \includegraphics[width=0.8\textwidth]{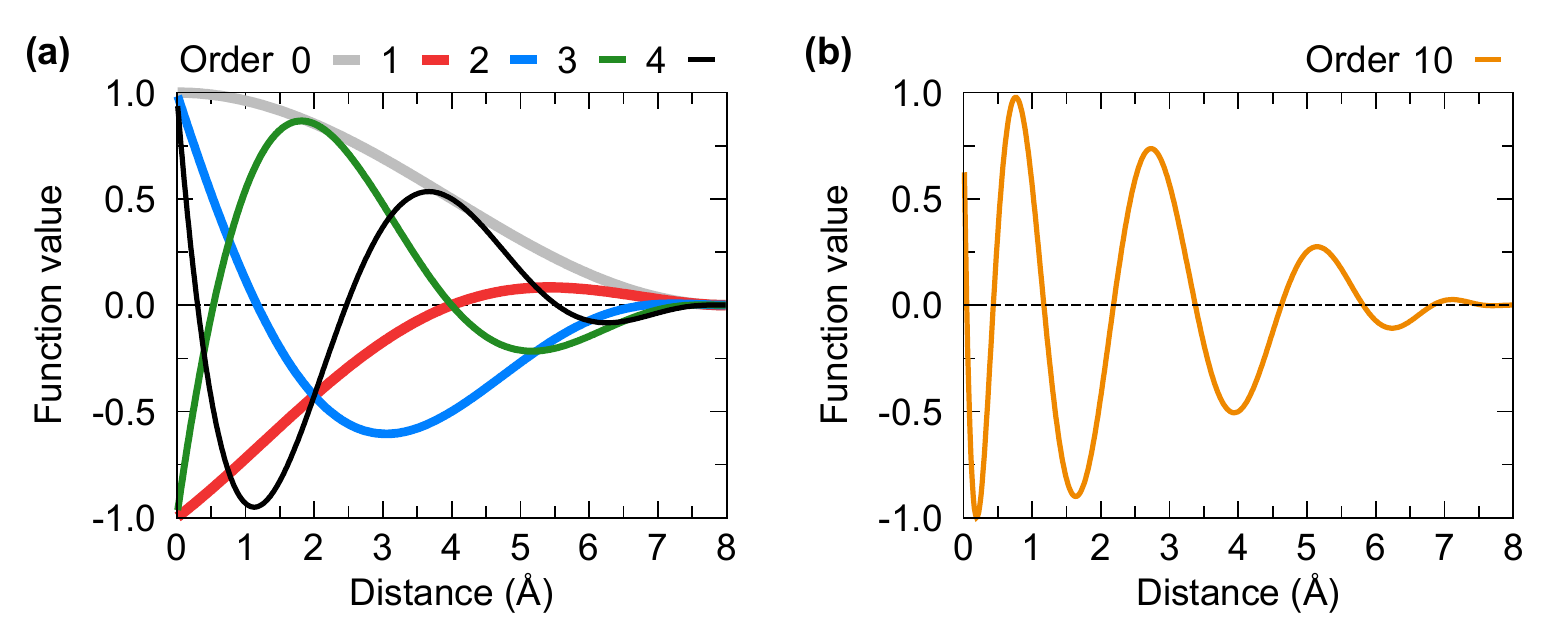}
  \caption{The Chebyshev descriptor (implemented in
    \ae{}net~\cite{artrith_implementation_2016, artrith_efficient_2017})
    enables the simulation of multicomponent compositions with many
    different chemical species. \textbf{(a)}~Product of the basis
    functions $\{\overline{\phi}_{\alpha}\}$ of
    equation~\eqref{eq:RDF-expansion} up to order~4 and the cosine
    cutoff function $f_c(R)$ for a radial cutoff of
    $R_{\textup{c}}=8$~\AA{}.  \textbf{(b)} shows the same for expansion
    order $10$.}
  \label{fig:chebyshev}
\end{figure*}

The original high-dimensional ANN potential by Behler and Parrinello (BP) introduced a set of \emph{symmetry functions} for the sampling of \emph{bond lengths} and \emph{bond angles} within local atomic environments~\cite{behler_generalized_2007, behler_atom-centered_2011}, which were in the spirit of earlier techniques for the symmetry-invariant representation of atomic coordinates~\cite{lorenz2004}.
Artrith, Urban, and Ceder (AUC) showed that the symmetry-function descriptor can be understood as a basis set expansion of the radial and angular distribution functions (RDF and ADF) and proposed to replace the original BP functions with an orthonormal basis set of Chebyshev polynomials~\cite{artrith_efficient_2017}.
The expansion of the RDF of the local atomic environment of atom $i$ can then be written as (the ADF expansion is analogous)
\begin{align}
\begin{aligned}
  &\textup{RDF}_{i}(R)
  \approx \sum_{\alpha=0}^{\alpha_{\textup{max}}^{\textsc{rdf}}}
          c_{\alpha}^{\textsc{rdf}}\phi_{\alpha}(R)
  \\
  &\text{with}\quad
  c_{\alpha}^{\textsc{rdf}}
  = \sum_{\vec{R}_{j} \in \sigma_{i}}
    \overline{\phi}_{\alpha}(R_{ij})f_{c}(R_{ij})
  \label{eq:RDF-expansion}
  \quad ,
\end{aligned}
\end{align}
where $\overline{\phi}_{\alpha}$ is the Chebyshev polynomial of order $\alpha$ and $\phi_{\alpha}$ is the orthonormal \emph{dual} function.
The cutoff function $f_{c}(R)=0.5[\cos(\pi\,R/R_{c}) + 1]$ for $R\leq{}R_{c}$ goes smoothly to zero at the cutoff radius $R_{c}$ and $R_{ij}=|\vec{R}_{j}-\vec{R}_{i}|$ is the distance of neighbor atom $j$ from the central atom $i$.
The RDF and ADF themselves already exhibit the symmetries of the potential energy, so that the joint sets of expansion coefficients $\{c_{\alpha}^{\textsc{rdf}}\} \cup \{c_{\alpha}^{\textsc{adf}}\}$ can be used as an invariant feature vector, and the precision and the length of the AUC feature vector can be adjusted by deciding on the polynomial orders $\alpha_{\textup{max}}^{\textsc{rdf}}$ and $\alpha_{\textup{max}}^{\textsc{adf}}$ at which the expansions are truncated.

\begin{figure*}[t]
  \centering
  \includegraphics[width=1.02\textwidth]{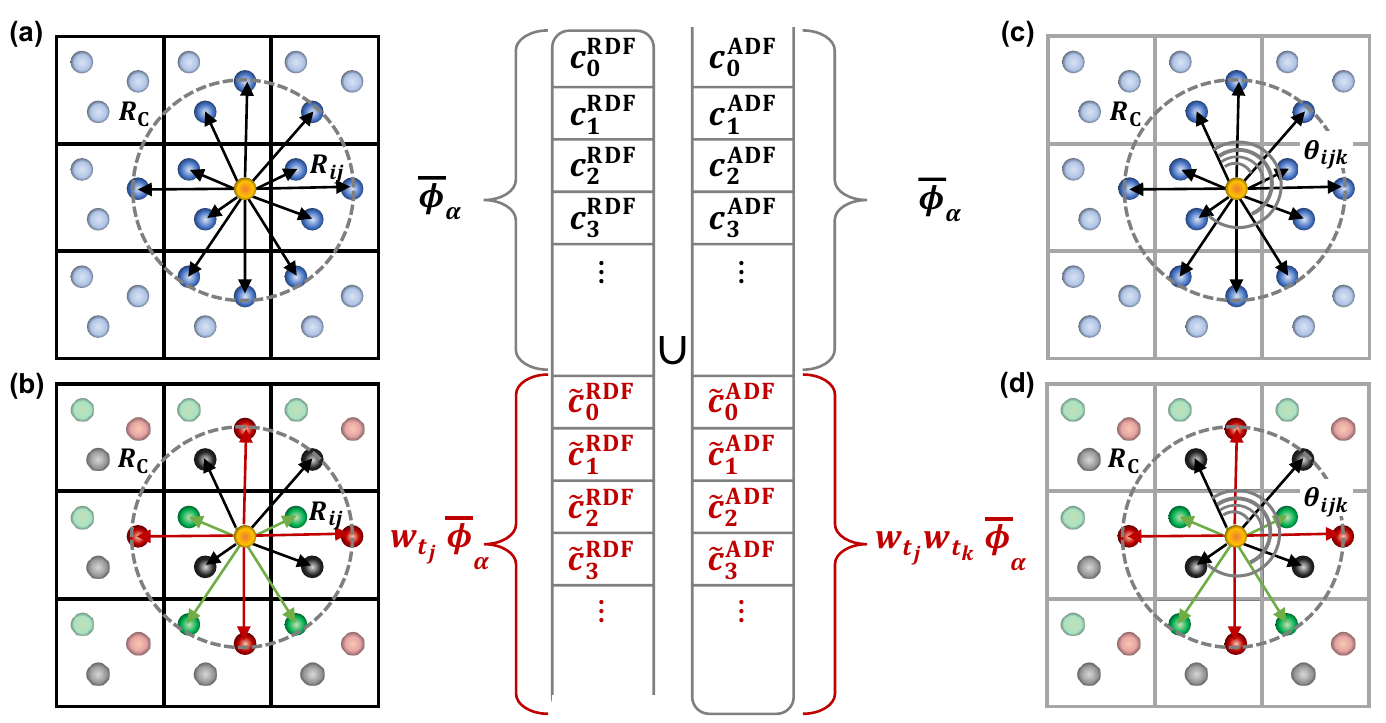}
  \caption{\label{fig:descriptor-radial-angular}%
    Schematic of the Artrith-Urban-Ceder (AUC) descriptor that is implemented in
    \ae{}net~\cite{artrith_implementation_2016}.  The atomic arrangement
    within the local atomic environment is described by the coefficients
    of an expansion in Chebyshev polynomials
    ($\overline{\phi}_{\alpha}$) of the \textbf{(a)}~radial and
    \textbf{(c)}~angular distribution functions. The atomic species
    (i.e., the local composition) is encoded in further sets of
    expansion coefficients, shown in \textbf{(b)} and \textbf{(d)}, obtained for basis functions that are
    weighted with species-specific weights $w_{t_{j}}$ ($t_{j}$ is the
    atom type of atom $j$) that can be calculated at no extra cost, see Eq. \eqref{eq:composition-descriptor}.}
\end{figure*}

\textbf{Figure~\ref{fig:chebyshev}} shows the Chebyshev basis functions, and \textbf{Figure~\ref{fig:descriptor-radial-angular}a~and~c} show a schematic of the construction of the feature vector for the local atomic \emph{structure}.

The expansion in equation~\eqref{eq:RDF-expansion} provides the featurization of the atomic positions within the local atomic environment, but it does not account for the different chemical species, since all atoms are considered equal.
To encode also information about atom types, an additional set of expansion coefficients is included as features, for which the contributions from each atom are \emph{weighted} with species-specific weights $w_{t_{j}}$ for atom type $t_{j}$~\cite{artrith_efficient_2017}
\begin{align}
\begin{aligned}
  \widetilde{c}_{\alpha}^{\textsc{rdf}}
  &= \sum_{\vec{R}_{j} \in \sigma_{i}}
    w_{t_j} \overline{\phi}_{\alpha}(R_{ij})f_{c}(R_{ij})
  \\
  \widetilde{c}_{\alpha}^{\textsc{adf}}
  &= \sum_{\vec{R}_{j} \in \sigma_{i}}
    \sum_{\vec{R}_{k} \in \sigma_{i}}
    w_{t_j} w_{t_k}
    \overline{\phi}_{\alpha}(\theta_{ijk})
    f_{c}(R_{ij})f_{c}(R_{ik})
  \quad ,
\end{aligned}
  \label{eq:composition-descriptor}
\end{align}
where $\theta_{ijk}$ is the cosine of the angle between atoms $i$, $j$, and $k$.
This weighted expansion gives rise to two more sets of expansion coefficients $\{\widetilde{c}_{\alpha}^{\textsc{rdf}}\}$ and $\{\widetilde{c}_{\alpha}^{\textsc{adf}}\}$ that can be calculated at essentially no additional cost together with the unweighted coefficients.
This featurization of the \emph{composition} is shown schematically in \textbf{Figure~\ref{fig:descriptor-radial-angular}b~and~d}.
Note that the length of the feature vector does not depend on the number of chemical species that are present in the atomic structure.

\subsubsection{Selecting the descriptor resolution}
\label{sec:descriptor-resolution}

The representation of the local atomic environment by expansion-based descriptors, such as the AUC descriptor discussed above, becomes more precise as the number of basis functions is increased.
However, the computational cost of both the featurization and the evaluation of the atomic energy ANNs also depends on the number of features.
The descriptor resolution should therefore be considered a hyperparameter that has to be optimized for a given application, so that the number of features is as large as needed and as small as possible.

For descriptors that are not based on expansion, an additional feature selection step can be introduced.
\emph{Feature engineering} and \emph{feature selection} are subject of current research, and various methods have been proposed for constructing descriptors and selecting relevant features automatically~\cite{cubuk_representations_2017, imbalzano_automatic_2018, jinnouchi_descriptors_2020, li_pairdistributionfunction_2020}.

\subsubsection{Selecting a radial cutoff}
\label{sec:radial-cutoff}

Another hyperparameter that should be optimized during the model construction phase is the cutoff radius $R_{c}$ of the local atomic environments.

The larger the cutoff radius is, the more information is available for featurization, and the more atomic structures can, in principle, be distinguished.
However, the computational cost of featurization also generally increases with the number of atoms within the local atomic environment, and the number of atoms scales as $R_{c}^{3}$.
For computational efficiency it is therefore beneficial to choose a radial cutoff that is as small as possible.

The radial cutoff $R_c$ can also strongly affect the convergence of training.
If $R_{c}$ is chosen too small, the feature vectors might contain insufficient information for the construction of accurate ANN potentials, resulting in poor generalization.
But if $R_{c}$ is chosen too large, the feature vectors might become dominated by irrelevant structural and compositional differences that do not actually affect the atomic energies, thus leading to poor training convergence.

Typical cutoff radii lie after the second or third coordination shell of the central atom, which corresponds to $\sim$6-8~\AA{} for metal oxides~\cite{artrith_neural_2013, artrith_implementation_2016, elias_elucidating_2016, artrith_efficient_2017} and $\sim$4--6~\AA{} for organic molecules~\cite{smith_ani-1_2017}.
It can be beneficial to increase $R_{c}$ further to capture also non-bonded interactions, such as hydrogen bonds, and for water cutoff distances of 6-10~\AA{} have been reported to give accurate results~\cite{morawietz_neural_2012,morawietz_how_2016,morawietz_interplay_2018}.

However, since the optimal cutoff range can be strongly dependent on the chemical system and application, it is necessary to perform a parameter study and compare the accuracy and transferability of the resulting ANNs in the validation stage of model construction (section~\ref{training}).

\subsection{Recipe: Artificial neural network model parameters}
\label{sec:ANN-parameters}

While the featurization determines the input dimension of the atomic energy ANNs, the internal architecture of the ANNs and the employed activation functions are also hyperparameters that need to be optimized.

\subsubsection{Selecting the ANN architecture}
\label{sec:architecture}

The \emph{architecture} of an atomic ANN is defined by the number of hidden layers and the number of neurons (nodes) in each hidden layer.
As discussed in section~\ref{sec:ANN-potentials}, the number of nodes in the input layer is defined by the dimension of the feature vector, and the output layer consists of a single node that returns the atomic energy.

The ANN architecture thus determines the model \emph{complexity}, in the sense that it determines the number of optimization parameters, i.e., the weights $\{w^{i}_{k,j}\}$ and $\{b_{i,j}\}$ of equation~\eqref{eq:artificial-neuron}.
If the ANN is too small, i.e., if it has too few hidden layers or nodes per layer, the MLP will not be flexible enough to reproduce the underlying PES well.
However, if the ANN is too large and too flexible, it might learn spurious information such as noise during training, which leads to overfitting (see section~\ref{training}).

It is possible to monitor for and to avoid overfitting even with large ANN architectures, such as deep ANNs with more than three layers, but larger-than-needed ANN architectures are also undesirable because the computational effort for training and ANN potential evaluation increases approximately as $N^{2}$ with the number of neurons per layer $N$.
Thus, generally the smallest ANN architecture yielding the required accuracy and transferability standards should be employed for constructing ANN potentials~\cite{artrith_high-dimensional_2012}.

Many reported accurate ANN potentials are based on architectures with two hidden layers.
The number of neurons per layer is a hyperparameter that has to be optimized for a given chemical system and application, though often $\sim$20--30~nodes per layer can already yield highly accurate ANN potentials even for complex materials~\cite{behler_first_2017}.
For large reference data sets and complex structure/composition spaces, architectures with more than 2~hidden layers and more than 50~nodes per layer may be beneficial~\cite{smith_ani-1_2017}.

\begin{figure*}[t]
  \centering
  \includegraphics[width=0.9\textwidth]{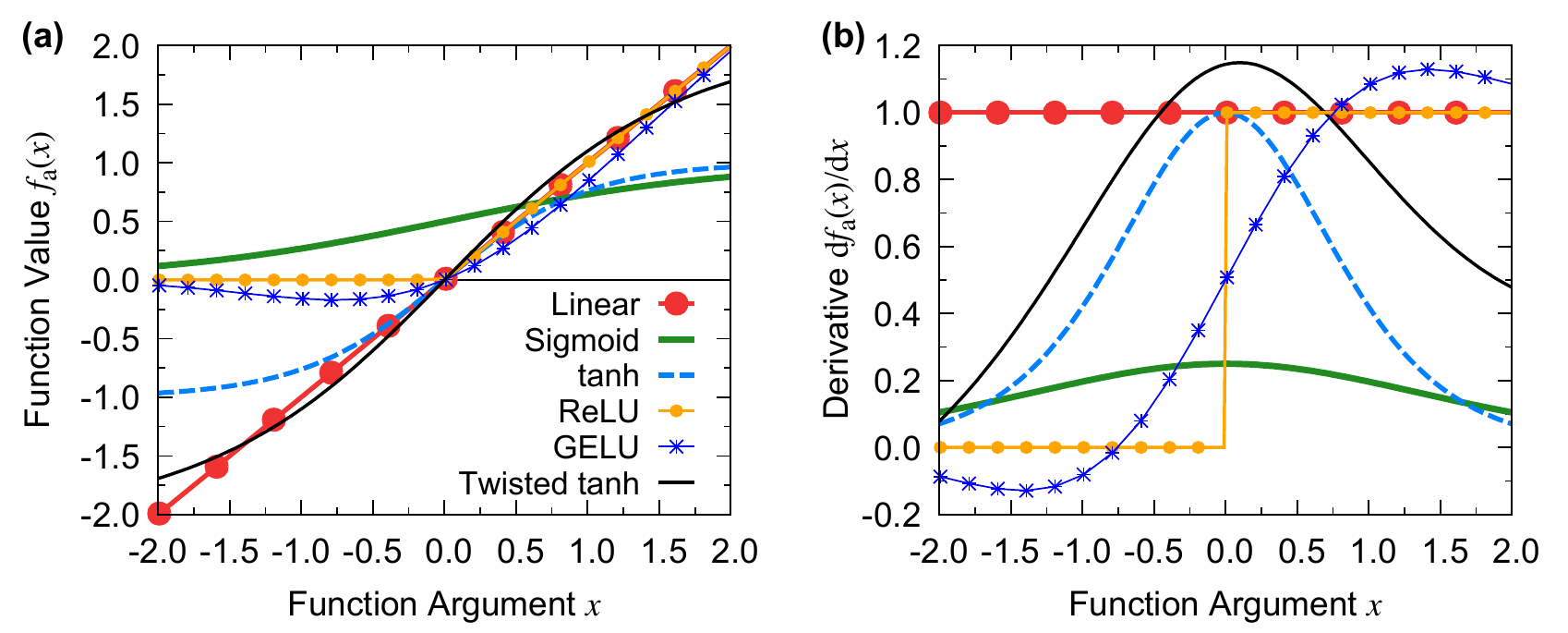}
  \caption{\label{fig:actfcts}%
    Plot of the six activation functions that are currently available
    in ænet~\cite{artrith_implementation_2016}. \textbf{(a)} Function
    values (output signals) for input values between -2 and
    +2. \textbf{(b)} Function derivatives for the same input values.}
\end{figure*}

\subsubsection{Selecting the activation functions}
\label{sec:activation-functions}

The choice of the activation function $f_{a}$ in equation~\eqref{eq:artificial-neuron} is another hyperparameter.
For ANN training with standard backpropagation methods (see section~\ref{training}), the activation function needs to be differentiable.
The activation function needs to be non-linear, or otherwise the ANN function becomes a linear model.
Further, it was found that monotonically increasing activation functions can accelerate the convergence of the weights during training\cite{wu09}.

Some of the most common activation functions and their derivatives used for atomic-energy ANNs are shown in \textbf{Figure~\ref{fig:actfcts}}.
The activation potential of biological neurons resembles a step function, and step-like or sigmoidal activation functions, such as the logistic function or the hyperbolic tangent, are also a popular choice for artificial neurons.
However, both functions are non-constant for only a small range of input values (activations), so that care must be taken during training to ensure a non-vanishing gradient of the \emph{loss function} (section~\ref{sec:loss_function}).
To avoid such saturation issues, a constant linear slope can be added to the sigmoidal function, an example of which is the \emph{twisted} hyperbolic tangent function~\cite{montavon_tricks_2012} shown in \textbf{Figure~\ref{fig:actfcts}}.
The rectified linear unit (ReLU) activation function has been introduced more recently~\cite{nair_rectified_2010} to avoid vanishing gradients in deep ANNs.
Although the derivative of the ReLU function exhibits a discontinuity at 0, good training results can be obtained in practice also for regression models such as ANN potentials~\cite{han_deep_2018, pattnaik_machine_2020}.
The Gaussian error linear unit (GELU) function~\cite{hendrycks_gaussian_2020} has similar properties to ReLU but does not exhibit any discontinuity in its derivative and is thus an even better choice for ANN potentials.
Many more activation functions have been proposed in the literature, and the development of activation functions is still an active area of research.

In conclusion, the activation functions for the hidden layers of ANN potentials have undergone an evolution over time, and the twisted hyperbolic tangent and the GELU function address the shortcomings of the previous generation and are typically good choices in practice.

Note that the range of output values that an artificial neuron can produce is determined by the co-domain of the activation function.
Therefore, the activation function of the final layer of atomic-energy ANNs is typically chosen to be the linear function, so that the output of the ANNs, i.e., the atomic energy, is unconstrained.

\subsection{Example: Model selection}
\label{sec:example-model-selection}

For the water example of section~\ref{sec:example-initial-data}, the hyperparameters were optimized during model construction.
The best compromise of accuracy on the validation set and computational efficiency was obtained for a model with a descriptor dimension of 51 (for hydrogen atoms) and 46 (for oxygen atoms), a radial cutoff of 6.35~\AA{}, two hidden layers, and 25~nodes per layer with hyperbolic tangent activation function~\cite{morawietz_interplay_2018}.

An example of an ANN potential for an inorganic solid material is the potential for amorphous LiSi alloys from reference~\citenum{artrith_constructing_2018}, for which the hyperparameters were also thoroughly optimized.
The final ANN potential employed an AUC descriptor with radial and angular expansion order of 10 ($=$~44 dimensions in total including the zeroth order), a cutoff radius of 8~\AA{}, two hidden layers with each 15 nodes, and hyperbolic tangent activation function.

\section{Model training and validation}
\label{training}

For a specific choice of hyperparameters (section~\ref{sec:model}), the ANN potential model needs to be \emph{trained} and \emph{validated} on different parts of the data set (section~\ref{sec:data}).
Model training is the process of optimizing the weight parameters $\{w^{i}_{k,j}\}$ and $\{b_{i,j}\}$ of Eq.~\eqref{eq:artificial-neuron} for all nodes of the ANN such that a \emph{loss} function $\mathcal{L}$ is minimized.
Since the optimization targets, i.e., the total energies of the reference structures, are given, this is a \emph{supervised learning} task.
Combining all weight parameters in a single set $W$, the weight optimization can be expressed as
\begin{align}
  W_{\textup{opt}}
  = \arg \min_{W} \mathcal{L}(W; S_{\textup{train}})
  \label{eq:loss-function-optimization}
  \quad ,
\end{align}
where $W_{\textup{opt}}$ is the set of optimal weight parameters and $S_{\textup{train}}$ is the \emph{training set} of all atomic structures from the reference data set that are used for training.
The remaining atomic structures make up the \emph{validation set} $S_{\textup{val}}$ that is used during training to monitor the progress, to detect overfitting, and to obtain an initial estimate of the ANN potential accuracy.
As indicated in the flow chart of \textbf{Figure~\ref{fig:TOC-figure}}, the model hyperparameters are typically varied until the model achieves optimal performance on the validation set.

\subsection{Recipe:}

To perform the optimization of equation~\eqref{eq:loss-function-optimization} in practice:
\begin{enumerate}
\item The data set needs to be split into training and validation sets,
\item The initial values for the weight parameters need to be set,
\item A suitable loss function needs to be defined, and
\item A training method has to be chosen.
\end{enumerate}

\subsubsection{Selecting training/validation sets}
\label{sec:validation-sets}

Both training set $S_{\textup{train}}$ and validation set $S_{\textup{val}}$ are derived from the overall reference data set.
The training\,:\,validation split (point~1.\ of the above list) is often between 90\%\,:\,10\% and 50\%\,:\,50\%, depending on the size of the reference data set.
The validation data should be selected randomly and should be representative for the entire reference data set.
The training and validation sets must not overlap, i.e., no atomic structure may be present within both sets.

The validation set is used only for obtaining an initial estimate of the ANN potential accuracy and its ability to generalize, but additional independent testing of the trained model is necessary (section~\ref{sec:testing-and-refinement}).
The main purpose of the validation set is to find the optimal set of hyperparameters that minimizes the validation error on unseen structures.

\subsubsection{Weight initialization and feature/target standardization}
\label{sec:feature_normalization}

The accuracy that a trained ANN model can achieve and the efficiency of solving the weight optimization of Eq.~\eqref{eq:loss-function-optimization} can strongly depend on the initial values of the weight parameters $W$ as well as the value range of the features and targets~\cite{thimm_high-order_1997}.
Feature/target normalization and the choice of initial weight parameters is therefore an important first step for the training of ANN potentials.

As discussed in section~\ref{sec:activation-functions} and shown in \textbf{Figure~\ref{fig:actfcts}b}, the gradient of typical activation functions is non-zero only for a narrow range of input values, which in turn depend on the value of the weights $w^{i}_{k,j}$ and $b_{i,j}$ and the magnitude of the features or node outputs $x_{i,j}$ (Eq.~\eqref{eq:artificial-neuron}).
If the initial weight parameters are chosen such that the activation function gradients are close to or identical to zero, standard methods for the weight optimization Eq.~\eqref{eq:loss-function-optimization} that follow the gradient of the loss function are inefficient.
This issue of vanishing gradients can be alleviated by ensuring that the output values of all neurons in each given layer are initially centered around zero~\cite{montavon_tricks_2012}.

A common approach is shifting and scaling the features (i.e., the descriptors of the local atomic environment of section~\ref{sec:featurization}) such that their values are centered around the point of inflection of the activation function (if applicable) and fall into the non-constant range of the activation function.
For example, for hyperbolic tangent activation functions (see section~\ref{sec:activation-functions}), the features can be shifted and scaled such that their variance is equal to~1 and the values are zero-centered~\cite{montavon_tricks_2012}.
Note that this standardization should be applied to each feature individually.
With this convention for feature standardization, the weight parameters should also be initialized such that the distribution of weights has a variance and center that is appropriate for the activation function used (e.g., a variance of 1 and center of 0 for the hyperbolic tangent).
In practice, initial weights are first drawn from a random distribution and then normalized.

The convergence of the learning process generally benefits from adjusting also the weights of the hidden layers such that the arguments of their nodes reside in the non-linear region of the activation function.
Various weight initialization methods have been developed and proposed in the literature, and an overview on common techniques can be found in reference~\citenum{thimm_high-order_1997} and the references therein.

\begin{figure*}
  \centering
  \includegraphics[width=0.80\textwidth]{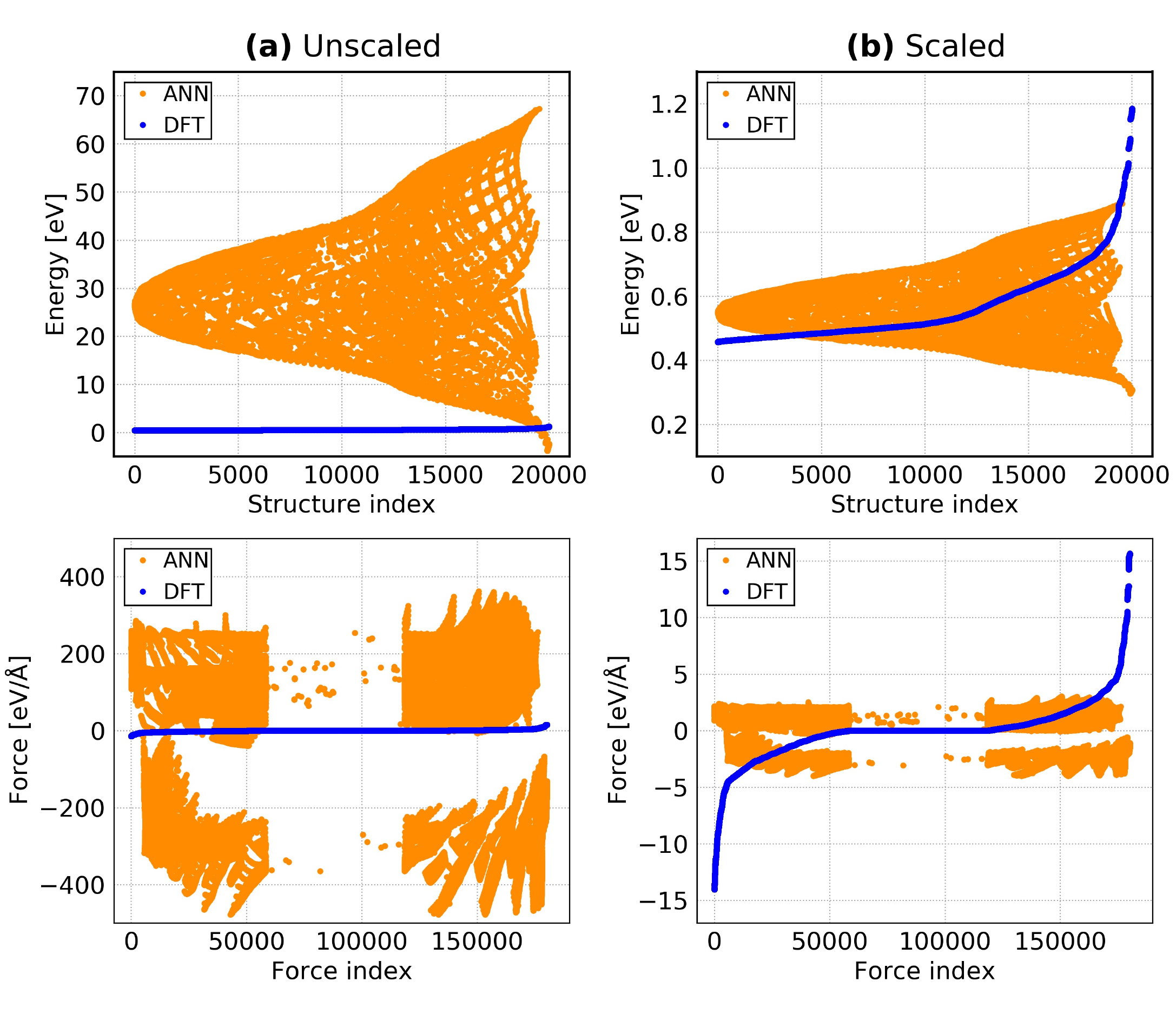}
  \caption{Distribution of ANN energies and forces before model training compared to their corresponding target values without (left panels) \textbf{(a)} and with (right panels) applying output normalization \textbf{(b)}. Closer alignment between initial ANN output and target values can be obtained by adjusting the outer ANN weights to match the mean and standard deviation of the target energy distribution, enabling faster model convergence with the number of training iterations. The data shown are reference configurations for a water monomer PES based on around ~20,000 configurations generated by a grid search~\cite{morawietz_entwicklung_2010}. The structures are sorted by ascending DFT reference energy and force.}
  \label{fig:outputnorm}
\end{figure*}

Although ANN potentials typically use a linear activation function in the output layer so that the energy is unconstrained, it is still beneficial to scale the target values so that large differences in the scale of the weights in the output layer to those in the other layers are avoided, and the initial ANN output is already close to the target values.
Thus, the structural energies are also commonly shifted and scaled such that they are zero-centered and have the same variance as the features.
Instead of standardizing the target values, the outer ANN weights can also be adjusted before model training so that the mean and standard deviation of the initial ANN output matches the target distribution (see \textbf{Figure~\ref{fig:outputnorm}}).

The impact of feature/target standardization and weight initialization on the initial state of an ANN potential is visualized in \textbf{Figure~\ref{fig:outputnorm}}.
Panel \textbf{(a)} of the figure shows that the initial predictions of the ANN potential before training can be orders of magnitude different from the DFT reference values, if the target energies and forces are not standardized.
After proper standardization, the initial predictions are of the same order of magnitude as the target values as shown in \textbf{Figure~\ref{fig:outputnorm}b}, which generally accelerates the training process significantly.

\subsubsection{Selecting a loss function}
\label{sec:loss_function}

The loss function $\mathcal{L}$ (sometimes also called \emph{objective function}) encodes the optimization targets, i.e., the properties that the ANN potential is trained to reproduce, such as the total structural energy and/or the interatomic forces.
Whether only energies or only forces should be included in the loss function depends on the planned application of the ANN potential, since the improved performance of the chosen target property usually causes the property that is left out to be of reduced accuracy.
On the other hand, training on both energies and forces simultaneously is computationally more demanding than training on each quantity separately.

\paragraph*{Energy training:} The most common choice of loss function $\mathcal{L}$ is for training on energy information only
\begin{align}
  \mathcal{L}_{E}(W)
  =& \frac{1}{N_{\textup{train}}}
     \sum_{\sigma\in{}S_{\textup{train}}}\!
       \frac{1}{2}\Bigl(
          E_{\textsc{ann}}(\sigma; W) - E_{\textup{ref}}(\sigma)
       \Bigr)^2
  \quad ,
  \label{eq:loss_energytrain}
\end{align}
where $N_\textup{train}$ is the number of atomic structures in the training set $S_{\textup{train}}$, $E_\textsc{ann}$ is the ANN potential energy of Eq.~\eqref{eq:E=sum-E_i}, $E_\textup{ref}$ is the corresponding reference energy, and the sum runs over all structures $\sigma$ within the training set.

Although some simulation techniques do not require gradients, such as Metropolis Monte-Carlo sampling~\cite{artrith_understanding_2014}, for most applications including geometry optimizations and MD simulations the interatomic forces need to be well represented.
Training on the energy-dependent loss function $\mathcal{L}_E$ of Eq.~\eqref{eq:loss_energytrain} can yield accurate forces, but a fine sampling of the phase space of interest might be required~\cite{witkoskie_neural_2005}.
Hence, a training set with a large number of structures might be needed, and the computational cost for the reference calculations can become limiting.
It is therefore often desirable to include the atomic forces as additional training targets~\cite{christensen_gradients_2020}.

\paragraph*{Force training:} In principle, it is possible to train an ANN potential on force information only by employing a loss function that is based on the force errors
\begin{align}
  \mathcal{L}_{F}(W)
  = \frac{1}{N_{\textup{train}}}
    \sum_{\sigma\in{}S_\textup{train}}
      \sum_i^{\textup{atoms}} \Bigl|
        \vec{F}_{i,\textsc{ann}}(\sigma; W) - \vec{F}_{i,\textup{ref}}(\sigma)
      \Bigr|^2
  \quad ,
  \label{eq:loss_forceonlytrain}
\end{align}
where $\vec{F}_{i,\textsc{ann}}(\sigma; W)$ is the force acting on the $i$-th atom in structure $\sigma$ as predicted by the ANN potential and $\vec{F}_{i,\textup{ref}}$ is the corresponding reference force.
For efficient force training, the ANN potential can be expressed such that the final layer returns an atomic force vector instead of an atomic energy~\cite{li_comparison_2018}.
However, the total structural energy can only be obtained up to an unknown constant from ANN potentials that predict forces, which makes it challenging to validate such potentials for the prediction of thermodynamic quantities.

\paragraph*{Energy and force training:} A more comprehensive loss function can be constructed by combining the energy and force loss functions from Eqs.~\eqref{eq:loss_energytrain} and~\eqref{eq:loss_forceonlytrain}, which can also be generalized to higher derivatives of the energy
\begin{align}
\begin{aligned}
  \mathcal{L}_{E,F}(W)
    = a_E\, \mathcal{L}_E(W)
    + a_F\, \mathcal{L}_F(W)
\end{aligned}
\label{eq:loss_directforcetrain}
\end{align}
where $a_E$ and $a_F$ are weights that determine the contributions of the energy and force errors to the overall loss function value.

The combined energy and force training ensures that information from both the potential energy and its gradient enter the training, which reduces the number of reference data points required for training~\cite{cooper_efficient_2020}.
However, since the forces are the negative gradient of the energy, training with the loss function of Eq.~\eqref{eq:loss_directforcetrain} requires the derivative of the Force with respect to the ANN weights (i.e., the second derivatives of the energy), which can become computationally demanding and is error-prone to implement in computer code.
We note that complex implementations can be avoided by utilizing efficient numerical schemes instead of fully analytical derivatives~\cite{smith_simple_2020}.

Another alternative is the translation of force information to additional energy reference data points by approximating the energy of atomic structures with displaced coordinates using a Taylor expansion~\cite{cooper_efficient_2020}.
This approach improves the force prediction accuracy by increasing the density of training points without requiring additional reference calculations.

\subsubsection{Training methods}
\label{sec:training-methods}

Various methods for ANN weight optimization (Eq.~\eqref{eq:loss-function-optimization}) have been proposed in the literature.
Most practical methods make use of the gradient of the loss function, i.e., the derivative of the loss function $\mathcal{L}$ with respect to the ANN weight parameters $W$, which can be efficiently calculated using error backpropagation~\cite{rumelhart_learning_1986, montavon_tricks_2012}.
The choice of training method ultimately depends on the size of the training set, the size of the structures in the training set (in terms of atoms), and the available computer hardware.

In general, two classes of training methods are distinguished, \emph{batch training} and \emph{online training} methods.
In batch training, the ANN weights are updated based on the actual value and gradient of the loss function, which requires evaluating the errors of all samples in the training set.
In constrast, in online training, ANN weights are updated sequentially based on the errors and gradients of each indiviual sample from the reference data set.
An intermediate approach is the online training with \emph{mini batches}, in which the weight updates are calculated based on blocks of data containing a specified number of samples.

An advantage of batch-training methods is that the second weight derivative (the Hessian) can be approximated more readily, such as in the limited-memory Broyden-Fletcher-Goldfarb-Shanno (BFGS) method~\cite{broyden_bfgs_1970, fletcher_bfgs_1970, goldfarb_bfgs_1970, shanno_bfgs_1970, liu_lbfgs_1989}, which can enable  convergence to solutions with lower residual loss function.
Batch training can also be parallelized trivially, since the errors of all atomic structures in the reference data set can be evaluated simultaneously.

One advantage of online training with stochastic gradient descent~\cite{montavon_tricks_2012} or related methods such as ADAM~\cite{kingma2014Adam} is the efficient evaluation of weight updates.
In addition, models trained with online methods often generalize well because of the regularizing effect of the approximate loss function evaluation~\cite{montavon_tricks_2012}.
In online training, the training \emph{schedule} can be controlled, i.e., the order in which samples are chosen from the reference data set, which can be useful for preferential training of atomic structures with high errors or with low energies (Boltzmann weighting).
More complex online-training methods, such as the extended Kalman filter~\cite{blank_ekf_1994, julier_ekf_2004}, can also make use of approximate Hessian information.

\subsubsection{Overfitting and extrapolation}
\label{sec:overfitting-and-extrapolation}

\begin{figure*}[t]
  \centering
  \includegraphics[width=1.0\textwidth]{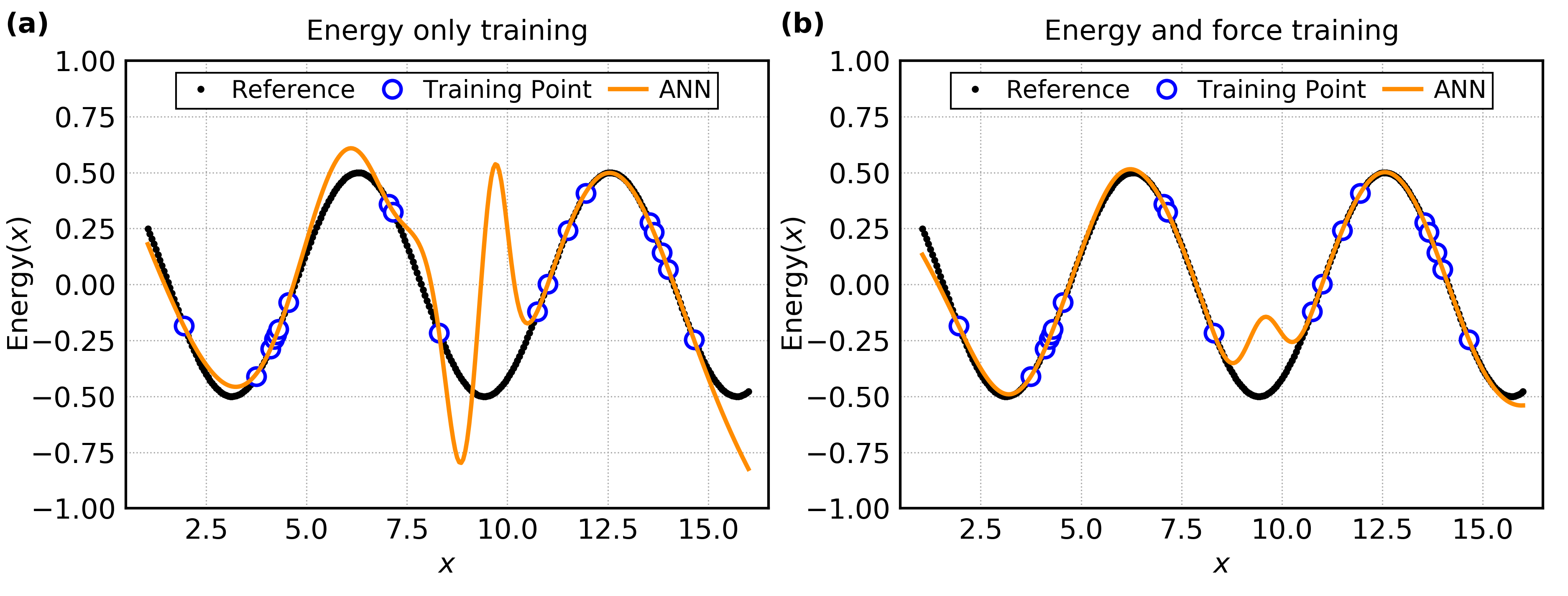}
  \caption{Overfitting and use of gradient information: Fitting of the simple model function $cos(x)$ in order to illustrate the effect of overfitting. Here, $x$ would correspond to an interatomic distance and the cosine function represents minima and maxima (or saddle points) of the potential energy surface.  When only energy information is used to optimize the artificial neural network (ANN) weights, the training points are accurately interpolated but poor results for points not in the training set are found \textbf{(a)}. Training the ANN to forces in addition to the reference energies, results in an improved representation of points not in the training set \textbf{(b)}.  For both examples, a $1-30-30-1$ ANN architecture was used, and the $x$ value was the only input.}
  \label{fig:overfitting}
\end{figure*}

ANN potential training using the methods of the previous section minimizes the loss function of Eq.~\eqref{eq:loss-function-optimization} for the training set only.
However, good performance for the training data does not necessarily imply that the potential will generalize to structures that were not included in the training process.

In order to estimate how well the MLP generalizes, the loss obtained for the validation set (section~\ref{sec:validation-sets}) is typically considered.
If the validation-set loss is similar to the training-set loss, it can be assumed that the ANN potential generalizes well for structures that are reasonably similar to those in the reference data set.
If the validation-set loss is significantly larger than the training-set loss, either \emph{overfitting} has occurred or the reference data set samples the structural space to sparsely so that \emph{extrapolation} is observed.

Overfitting is the phenomenon when the ANN function reproduces the training samples with high accuracy but at the cost of introducing unreasonable behavior in the regions between the training points (see \textbf{Figure~\ref{fig:overfitting}a}).
In the case of noisy reference data, overfitting also means that the ANN incorrectly reproduces the noise and not only the expected value of the targets.
Overfitting occurs when the ANN is too flexible for the size and density of the training set, i.e., if the training set contains too few data points or if the model complexity is too great and an adjustment of the hyperparameters (see section~\ref{sec:model}) is needed.
Overfitting can be reduced by introducing regularization terms in the loss function~\cite{hoerl_ridge_regression_1970, montavon_tricks_2012} or by extending the training set, for example, by including force information (see \textbf{Figure~\ref{fig:overfitting}b}).
During training, overfitting can be detected by monitoring the training-set and validation-set loss, which start to diverge at the onset of overfitting.

As discussed in the data selection section~\ref{sec:data}, ANNs are unreliable for extrapolation~\cite{haley_extrapolation_1992}.
If the training set samples the structure space too coarsely, regions of the potential energy surface may be insufficiently represented.
An example is shown in \textbf{Figure~\ref{fig:overfitting}b}.
Underrepresented regions can also be identified by the validation set, if at least some relevant structures are present.

\begin{figure*}[t]
    \centering
 \includegraphics[width=1.1\textwidth]{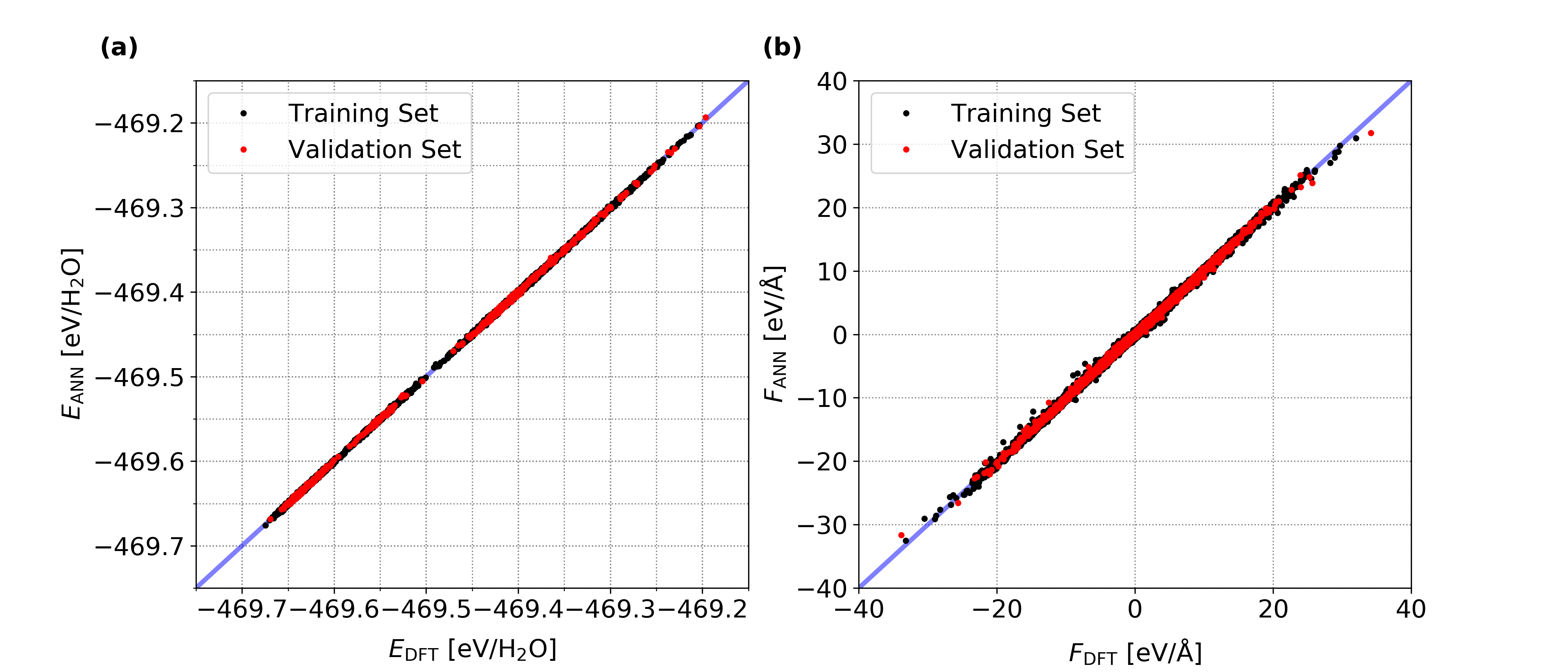}
 \caption{Evaluation of the training set (black circles) and validation set (red circles) accuracy of the machine learning potential (MLP) for liquid water trained on the initial reference data described in section~\ref{sec:data}.  \textbf{(a)}~ANN energies and \textbf{(b)}~the corresponding atomic forces plotted against their reference DFT values.}
 \label{fig:diagonal-plots}
\end{figure*}

\subsection{Example: Training and validation}
\label{sec:training-validation-example}

Returning to our example case on liquid water, \textbf{Figure~\ref{fig:diagonal-plots}} examines the training and validation set accuracy of the MLP trained on the initial data set described in section~\ref{sec:recipe-initial-data} based on a training\,:\,validation split of 90\%\,:\,10\%.
It can be seen that both training and validation set are accurately represented by the initial MLP across the full range of total energy and atomic force values with low validation root mean squared errors (RMSEs) (1.4~meV/\ce{H2O} and 83.2~meV\AA{}, respectively) that are on the same level as the corresponding training values (1.2~meV/\ce{H2O} and 84.7~meV/\AA{}).

These low figures and the close correspondence between training and validation errors indicate that the initial MLP has been sufficiently optimized and a close to optimal choice of hyperparameters has been found.
There are no obvious outliers and the accuracy is similar over the entire energy range, i.e., no energy region is underrepresented.
Now, the model can be tested in the intended application to quantify its accuracy.

\section{Model testing and active learning}
\label{sec:testing-and-refinement}

\begin{figure*}
    \centering
    \includegraphics[width=0.8\textwidth]{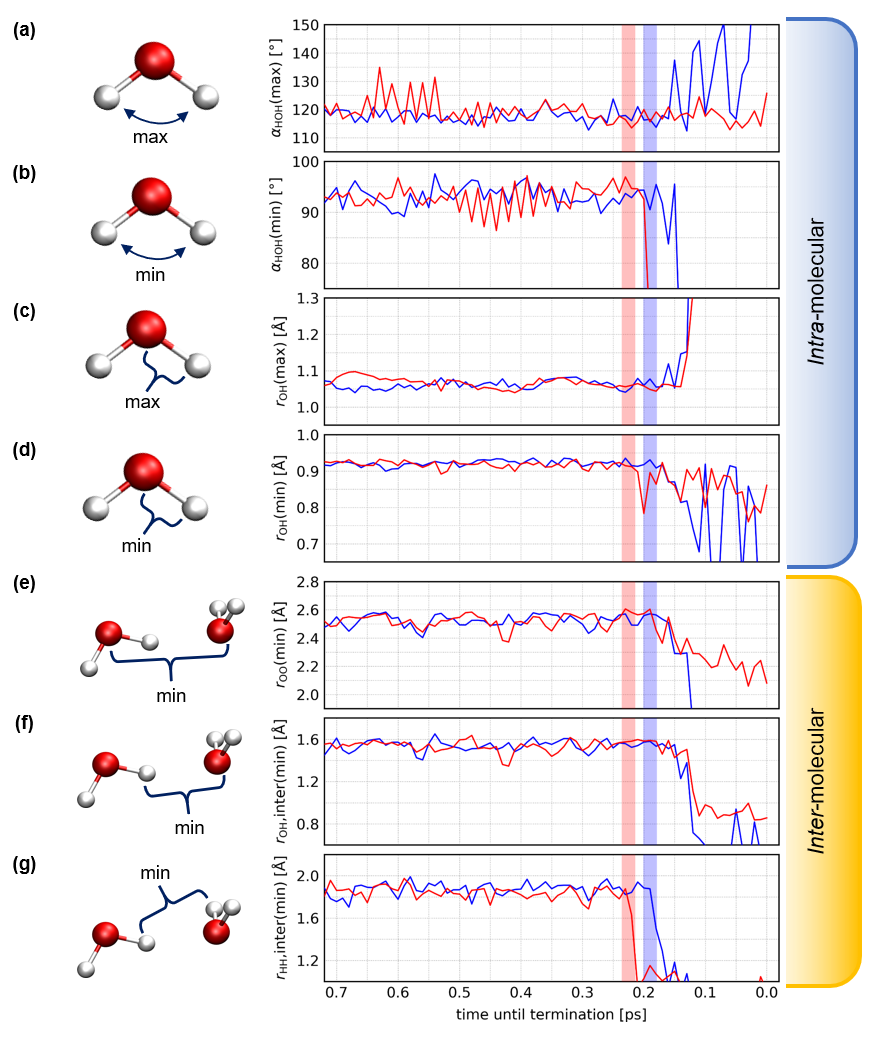}
    \caption{Monitoring of critical interactions for two molecular dynamics (MD) simulations (shown in red and blue) performed with the initial liquid water MLP. Both simulations ended prematurely due to instabilities in the potential. The x-axis shows the remaining simulation time until the end of the simulation. In panels \textbf{(a)}~and~\textbf{(b)} the intramolecular water angle with the shortest and largest value, respectively, across all angles in the simulation cell is shown for each time step. Correspondingly, in panels \textbf{(c-g)}~extrema of different intra- and inter-molecular distances are displayed. Instabilities are first observed in the minimum distance between inter-molecular hydrogen atoms (marked by the shades area) indicating that this interaction type might not be properly represented in the initial MLP even though it accurately represents all reference configurations.}
    \label{fig:critical_interactions}
\end{figure*}

Once an ANN potential has been obtained that performs well on the validation set (section~\ref{training}), the potential needs to be tested in actual applications, such as MD simulations.
Generally, an initial data set generated as described in section~\ref{sec:data}, e.g., by sampling from MD simulations and/or by manually perturbing equilibrium structures, does not ensure that the configuration space visited during the intended application is fully covered and can therefore lead to ANN potentials that exhibit instabilities in simulations.
This is exemplified in our water case study when the initial MLP was first employed in MD simulations.
As demonstrated in \textbf{Figure~\ref{fig:critical_interactions}}, the initial MLP exhibits stability issues leading to a premature termination of the simulation even though its validation accuracy is high (\textbf{Figure~\ref{fig:diagonal-plots}}) and its initial training set comprises of a diverse set of structures which cover a broad energy and force range (\textbf{Figure~\ref{fig:water-initial-data}}).
A detailed analysis of the corresponding trajectories reveals the underlying reason for the stability issue: the interactions between inter-molecular hydrogen atoms is not properly described.

Such generalization issues can be addressed by iteratively including additional data points in the reference data set, as shown in the outer loop of \textbf{Figure~\ref{fig:TOC-figure}}.
Here, a model trained on an initial data set is used in preliminary applications and then \emph{retrained} once it encounters configurations for which the model prediction shows a low accuracy.
The challenge here is how to identify such configurations with a high model uncertainty.

While in principle one could recompute all configurations generated with a given model by the underlying reference method to detect inaccurately described structures, this is computationally very inefficient.
A solution to this is to make use of \emph{active learning} approaches that select the next training set iteration \emph{from unlabeled data} in an automated fashion with the benefit that the amount of expensive reference calculations is limited.

Active learning approaches for the construction of MLPs \cite{artrith_high-dimensional_2012, lookman_active_2019, jinnouchi_onthefly_2020} comprise of three steps which are executed in a loop until the desired model performance is reached:
\begin{enumerate}
\item Efficient \emph{exploration} of the configuration space,
\item \emph{Selection} of relevant configurations and labeling, i.e. calculations of reference energies and forces,
\item Followed by model \emph{retraining}.
\end{enumerate}
One can distinguish between on-the-fly active learning \cite{podryabinkin_active_2017, jinnouchi_onthefly_2020, Vandermause_2020} in which retraining happens \emph{during} the simulation and offline active learning where the next iteration of training structures is first accumulated, then a new model is trained, and subsequently a new simulation with the improved MLP will be launched.

The crucial step in an active learning approach is the selection of data points with high model uncertainty without knowing their reference properties beforehand.
The goal is to find a query strategy to decide if a given configuration is already well described by the current ANN potential or if it should be added to the reference data set.
Approaches proposed in the literature generally belong to one of the following classes:

\paragraph{Data set reduction approaches:} A large set of candidate configurations is generated with some sampling strategy, redundant/similar configurations are removed, and additional reference calculations are only performed for the configurations that are most different from those in the present reference data set.
For example, Bernstein et al.\ selected a subset of configurations from relaxation trajectories by (1)~Leverage-score CUR algorithm, and (2)~Flat histogram sampling (selection from low-density regions) with Boltzmann-probability bias~\cite{bernstein_novo_2019}.

\paragraph{Query-by-committee approaches:} An ensemble of models is used to evaluate a set of candidate configurations that were generated with some sampling strategy using one specific model.
The standard deviation of the energy (and potentially the forces) across the committee of models is then used as an uncertainty estimate~\cite{artrith_high-dimensional_2012, smith_less_2018, schran_committee_2020}.
Ensemble methods can be also combined with data set reduction techniques as described above~\cite{smith_less_2018}.

\paragraph{Statistical uncertainty quantification:} Statistical inference, e.g., based on the Bayesian framework~\cite{BoxTiao2011BayesianInference}, can be used to estimate the uncertainty of predictions.
GPR models provide an intrinsic uncertainty estimate that can be employed~\cite{Vandermause_2020, jinnouchi_onthefly_2020}.
In ANN potentials, \emph{dropout} can provide an uncertainty estimate~\cite{wen2020dropout}, i.e., the \emph{query-by-committee} can be build into the ANN architecture.
In the case of MLP methods that depend linearly on the model parameters, such as moment tensor potentials~\cite{podryabinkin_active_2017}, extrapolation can also be detected on-the-fly, which can be exploited for active learning.

\subsection{Example: Model testing and refinement}
\label{sec:example-testing-refinement}

\begin{figure*}
    \centering
    \includegraphics[width=1.0\textwidth]{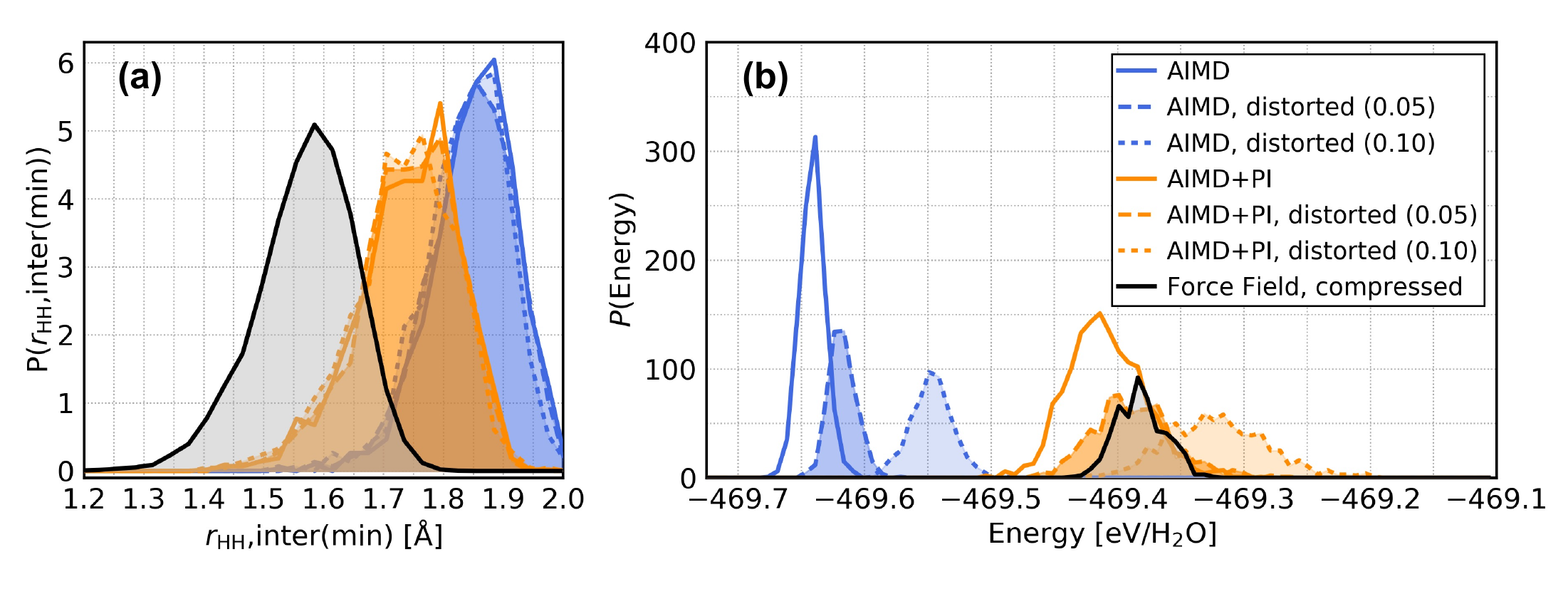}
    \caption{Iterative refinement of the initial liquid water dataset with additional structures from compressed force field MD simulations: \textbf{(a)} Distribution of the minimum inter-molecular hydrogen distance across the configurations in the reference data set (blue and orange) and the high density configurations obtained form force field MD (grey). \textbf{(b)} Energy distribution of the initial structures and the added compressed structures which together form the second iteration of the reference data set.}
    \label{fig:energydist}
\end{figure*}

In the case of the liquid water MLP, the initial dataset was extended manually instead of using an active learning technique since the underlying reason for its stability issue could be identified by the analysis summarized in \textbf{Figure~\ref{fig:critical_interactions}}.
To improve the representation of the ill-described inter-molecular hydrogen interactions additional reference structures were obtained from compressed MD simulations (employing an artificially increased density and a higher temperature) with a classical force field.
As shown in \textbf{Figure~\ref{fig:energydist}}, these structures sample short hydrogen-hydrogen distances and are located in a high-energy region.
After labeling the additional structures with the corresponding DFT energies and forces, the retrained MLP could be applied in MD simulations without any stability issues even at elevated temperatures.
Additional rounds of model testing and refinement at higher temperatures followed and the final reference set comprising of a total of 9,189 structures can be obtained online from reference~\citenum{chen_aenetlammps_2020}.

\section{Discussion and Outlook}
\label{sec:discussion}

In this tutorial review, we outlined common strategies for the construction of ANN-based machine-learning potentials.
However, even if the best practices for the construction of MLPs are followed and active learning approaches are implemented for data selection, there are still remaining challenges that hinder the universal application of MLPs.

Depending on the size of the relevant configuration space, one significant challenge can be the expense of the reference calculations.
Compiling a reference data set that captures all regions of the potential energy surface of a multi-component system can be a formidable challenge.

The construction of MLPs can be simplified by focusing on (1)~the description of specific parts of the potential-energy, (2)~certain regions of the system, or (3)~using MLPs alongside full first-principles calculations, instead of describing the full PES of all atoms in the simulation box with a single MLP.
Compared to general MLPs these \emph{specialized MLP} approaches are generally more robust, require a lower number of expensive high-level reference calculations, and are easier to converge during training, while having the downside of being computationally more involved during the execution of the actual simulation.

Potentials fitted not on the full potential energy surface but on the differences to some reference are commonly referred to as \emph{delta ML approaches}.

\paragraph*{Energy decomposition approaches:} Rather than learning the full potential energy, long-range contributions such as electrostatics or van der Waals (vdW) interactions can be removed before model training and the remaining short-range energy is used as the target property which can be more easily described by atomic ANNs that depend on local environments.
\begin{align}
  E^{\textup{total}}
  = E_{\textsc{ann}}^{\textup{short}} + E_{\textsc{ann}}^{\textup{elec}} + E^{\textup{vdW}}
  \label{eq:vdw-ANN-potential}
  \quad ,
\end{align}
The removed energy contributions can then be added back in by employing expressions that explicitly consider the physical distance dependence.
This can be done by either employing already available analytic expression, as in the case of vdW interactions~\cite{morawietz_densityfunctional_2013, yao_tensormol0_2018} (e.g. with Grimme's D2/D3 methods~\cite{grimme_semiempirical_2006,grimme_consistent_2010}), or by training separate ML models, for example to represent atomic charges for calculating long-range electrostatics based on Coulomb's law~\cite{artrith_high-dimensional_2011,morawietz_neural_2012, yao_tensormol0_2018}.
A dependence of the MLP energy on long-ranged features can also be directly included, avoiding the need to explicitly model atomic charges which are no uniquely defined observables~\cite{grisafi_incorporating_2019}.

\paragraph*{Energy-difference approaches:} Another group of delta ML approaches focuses on the prediction of energy differences between two reference methods of different quality.
Here the energy difference between a lower-level method and a higher-level quantum-mechanics (QM) based method is predicted by an MLP.
An early example of such a composite strategy in which an ML correction is added to a computationally efficient but less accurate QM method is the delta-machine learning approach by Ramakrishnan et al.~\cite{ramakrishnan_big_2015}.
Other examples using different levels of theory are discussed in references \citenum{sun_fast_2019}~and~\citenum{ramakrishnan_big_2015}.
Instead of predicting \emph{total energy differences}, molecular-orbital-based schemes model high-level electronic structure correlation energies using inputs from Hartree–Fock calculations with the goal of being more transferable across chemical systems \cite{welborn_transferability_2018, cheng_universal_2019}.

\paragraph*{Embedding approaches:} Following the spirit of QM/MM approaches~\cite{honig_implications_1971, warshel_theoretical_1976, aqvist_simulation_1993, mulholland_ab-initio_2000, senn_current_2007, magalhaes_modelling_2020}, delta MLPs can be developed to focus on certain regions in space within the entire system.
Those regions of the system that are not described by the MLP are instead described by molecular mechanics-based or lower-level QM methods~\cite{zhang_qmmm_2018}.
This strategy can also be combined with energy-difference approaches, for example by predicting the energy difference between a low-level QM method and a high-level QM method by an MLP for atoms inside a limited spatial region which is coupled to molecular mechanics-based interactions for describing the larger environment (making it a QM/MM/ML approach in which the \emph{delta} refers to energy difference \emph{and} spatial separation)~\cite{shen_molecular_2018}.

\paragraph*{Domain-limited approaches:} Finally, \emph{specialized} MLPs can be employed alongside conventional first principles calculations to accelerate the calculation of QM properties.
These potentials are often focused on a specific region of the configurational space and generally do not need to be trained to the highest degree of accuracy since final first principles reference calculations are included as part of the full workflow.
Examples are the use of ML-accelerated geometry optimizations in which initial structures are pre-optimized with an MLP, followed by a final ab initio optimization that requires fewer steps to convergence since its input structure is already close to the ground state~\cite{peterson_acceleration_2016, jacobsen_on-the-fly_2018}.
Specialized MLPs have also been employed in combination with evolutionary algorithms to determine the phase diagram of amorphous alloys~\cite{artrith_constructing_2018}.

\section{Summary}
\label{sec:summary}

Machine-learning interatomic potentials enable accurate and efficient atomic-scale simulations of complex systems that are not accessible by conventional first principles methods, but for many systems of interest machine-learning potentials have not yet been developed.
Here, we reviewed common strategies and best practices for the construction of machine-learning potentials based on artificial neural networks (ANN).
Data selection, model selection, training/validation, and testing/refinement have been exemplified using practical examples.
The number of available tools and data sets for machine-learning potential applications is rapidly growing, and we refer the reader to a curated and editable list at \url{https://github.com/atomisticnet/tools-and-data} with a collection of free and open-source tools and data resources.
As discussed, the construction of ANN potentials is still a complex and manual process involving many steps.
Recipes are provided here with the hope that in future more automated and standardized workflows for ANN construction will be established, so that the method can achieve its full potential in accelerating the prediction of materials and molecular properties with an unparalleled combination of accuracy and speed.

\section*{Acknowledgments}
\noindent
N.A.\ acknowledges support by the Columbia Center for Computational Electrochemistry (CCCE).
N.A.\ and A.U.\ thank the Extreme Science and Engineering Discovery Environment (XSEDE), which is supported by the National Science Foundation under Grant No. ACI-1053575, for supporting the development of the atomic energy network (ænet) package.
A.M.M.\ and J.K.\ acknowledge support by the state of Baden-Württemberg through bwHPC and the German Research Foundation (DFG) through grant no INST 40/575-1 FUGG (JUSTUS 2 cluster).
A.M.M.\ and J.K.\ also thank the Deutsche Forschungsgemeinschaft (DFG, German Research Foundation) for supporting this work by funding -- EXC2075 -- 390740016 under Germany's Excellence Strategy and acknowledge the support by the Stuttgart Center for Simulation Science (SimTech).
T.M.\ thanks T.E.\ Markland and O.\ Marsalek for fruitful discussions.

\section*{Competing interests}
\noindent
The authors declare no competing financial interests.

\section*{Code details}
\noindent
Companion materials can be accessed at \url{https://github.com/atomisticnet/MLP-beginners-guide}, which includes a Jupyter notebook demonstrating the construction of an ANN potential and a second notebook that allows reproducing the failure analysis of \textbf{Figure~\ref{fig:critical_interactions}}.

\end{document}